\documentclass[12pt,journal,final,onecolumn]{IEEEtran}

\usepackage[dvips]{graphicx}
\usepackage{epsfig}
\usepackage[cmex10]{amsmath}
\usepackage{amssymb}
\usepackage{amsthm}
\usepackage{amsfonts}
\usepackage{bm}
\usepackage{cite}
\usepackage[tight,footnotesize]{subfigure}

\graphicspath{{figs/}}

\newcommand{\mc}[1]{\mathcal{#1}}

\newcommand{\msf}[1]{\mathsf{#1}}

\newcommand{\SINR}{\msf{SINR}}
\newcommand{\SNR}{\msf{SNR}}

\newcommand{\defeq}{\triangleq}
\newcommand{\Pp}{\mathbb{P}}
\newcommand{\E}{\mathbb{E}}
\newcommand{\N}{\mathbb{N}}
\newcommand{\Z}{\mathbb{Z}}

\newcommand{\R}{\mathbb{R}}

\newcommand{\C}{\mathbb{C}}

\newcommand{\ind}{{1\hspace*{-0.8ex}{1}}}

\newcommand{\abs}[1]{\lvert{#1}\rvert}
\newcommand{\card}[1]{\abs{#1}}
\newcommand{\norm}[1]{\lVert{#1}\rVert}
\newcommand{\iid}{i.\@i.\@d.\ }
\newcommand{\T}{\msf{T}}

\DeclareMathOperator{\tr}{tr}

\DeclareMathOperator{\diag}{diag}

\DeclareMathOperator{\Ei}{Ei}

\newtheorem{lemma}{Lemma}

\newtheorem{theorem}[lemma]{Theorem}

\newtheorem{innercustomtheorem}{Theorem}
{\renewcommand\theinnercustomtheorem{#1}\innercustomtheorem}%
{\endinnercustomtheorem}

\theoremstyle{definition}
\newtheorem*{definition}{Definition}

\newtheorem{egdummy}{Example}
{%
    \begin{egdummy}[#1]%
    \upshape%
}%
{%
    \qed%
    \end{egdummy}%
}

\newtheoremstyle{myremark}%
{\topsep}{\topsep}{\normalfont}{\parindent}{\itshape}{:}{ }{}

\theoremstyle{myremark}
\newtheorem*{remark}{Remark}

\newcommand\shortintertext[1]{%
\ifvmode\else\\\@empty\fi
\noalign{%
\penalty0%
\vbox{\mathstrut}%
\penalty10000%
\vskip-\baselineskip
\penalty10000%
\vbox to 0pt{%
\normalbaselines
\ifdim\linewidth=\columnwidth
\else
\parshape\@ne
\@totalleftmargin\linewidth
\fi
\vss
\noindent#1\par}%
\penalty10000%
\vskip-\baselineskip}%
\penalty10000}

\begin{document}

\title{Computation Alignment: Capacity Approximation without Noise Accumulation}

\author{Urs Niesen, Bobak Nazer, and Phil Whiting%
\thanks{U.~Niesen and P.~Whiting are with the
Mathematics of Networks and Communications Research Department, Bell Labs, 
Alcatel-Lucent. B.~Nazer is with the ECE Department, Boston University.
Emails: urs.niesen@alcatel-lucent.com, bobak@bu.edu, pwhiting@research.bell-labs.com.}%
\thanks{This paper was presented in part at the Allerton Conference
on Communication, Control, and Computing, September 2011.}%
\thanks{This work was supported in part by AFOSR  under grant FA9550-09-1-0317.}%
}

\maketitle

\begin{abstract} 
    Consider several source nodes communicating across a wireless
    network to a destination node with the help of several layers of
    relay nodes. Recent work by Avestimehr \emph{et al.} has
    approximated the capacity of this network up to an additive gap. The
    communication scheme achieving this capacity approximation is based
    on compress-and-forward, resulting in noise accumulation as the
    messages traverse the network. As a consequence, the approximation
    gap increases linearly with the network depth. 

    This paper develops a \emph{computation alignment} strategy that
    can approach the capacity of a class of layered, time-varying
    wireless relay networks up to an approximation gap that is
    independent of the network depth. This strategy is based on the
    compute-and-forward framework, which enables relays to decode
    deterministic functions of the transmitted messages. Alone,
    compute-and-forward is insufficient to approach the capacity as it
    incurs a penalty for approximating the wireless channel with
    complex-valued coefficients by a channel with integer coefficients.
    Here, this penalty is circumvented by carefully matching channel
    realizations across time slots to create integer-valued effective
    channels that are well-suited to compute-and-forward. Unlike prior
    constant gap results, the approximation gap obtained in this paper
    also depends closely on the fading statistics, which are assumed to
    be i.i.d. Rayleigh.
\end{abstract}

\section{Introduction}
\label{sec:intro}

Consider a line network, consisting of a single source communicating to
a single destination via a sequence of relays connected by
point-to-point channels. The capacity of this simple relay network is
achieved by decode-and-forward and is determined solely by the weakest
of the point-to-point channels. As a consequence, the performance of the
optimal scheme is unaffected by noise accumulation, regardless of the
length of the relay network. This raises the question whether the same
holds true in general multi-user wireless relay networks, i.e., if the
capacity depends on the network depth. In this paper, we investigate
this question in the context of multiple sources communicating with a
single destination across a multi-layer wireless relay network.

\subsection{Motivation and Summary of Results}
\label{sec:intro_motivation}

In a multi-layer wireless relay network, each relay observes a noisy
linear combination of the signals transmitted by the relays in the
previous layer. In order to avoid noise accumulation, the relays should
perform some type of decoding to eliminate noise at each layer. A
natural approach is to use decode-and-forward, in which  each layer of
relays decodes the messages sent by the previous layer and retransmits
them, just as in the line network mentioned above. Unfortunately, while
the performance of this scheme is independent of the network depth, it
is often interference limited and, as a result, its performance can
diverge significantly from the capacity. 

Instead of combating interference, as is done in the decode-and-forward
approach, other communication strategies embrace the signal interactions
introduced by the wireless channel. One such strategy is
compress-and-forward, in which each relay transmits a compressed version
of its received signal. Such strategies can offer significant advantages
over decode-and-forward. Indeed, recent work by Avestimehr \emph{et
al.}~\cite{avestimehr11} has shown that, for a large class of wireless
relay networks that includes the layered network model considered here,
compress-and-forward approximately achieves capacity up to a gap
independent of the power constraints at the nodes in the network. 

One important feature of this approximation guarantee is that it is
uniform in the channel coefficients and hence the fading statistics.
However, since the compress-and-forward scheme does not remove noise at
each relay, noise accumulates from one layer in the network to the next.
As a consequence, the approximation gap in~\cite{avestimehr11} (and
related ones such as those based on noisy network coding \cite{lkec11})
increases linearly with the number of layers in the relay network. Thus,
as the depth of the network increases, the approximation guarantee
becomes weaker.  

In this paper, we make progress on this issue by deriving a new capacity
approximation result for the time-varying, multi-layer relay network
with an approximation gap that is independent of the depth of the
network. However, unlike the approximation result in
\cite{avestimehr11}, our guarantee depends on the fading statistics.
Specifically, we assume that each channel coefficient is drawn
independently according to a Rayleigh distribution. 

Our approach is built around the compute-and-forward framework proposed
by \cite{nazer11a}. In this framework, each transmitter encodes its
message into a codeword drawn from the same lattice codebook. As a
result, all integer combinations of codewords are themselves codewords,
enabling relays to decode linear functions of the transmitted codewords
rather than treating interference as noise. If these functions are
invertible, then the destination can use them to infer its desired
messages. 

While the use of lattice codes seems like a natural fit for this
setting, it alone is insufficient to approach the network capacity, as
was shown recently in \cite{niesen12}. The primary reason is that this
scheme approximates the wireless channel with complex-valued channel
gains by a channel with integer-valued channel gains. The residual
signals not captured by this integer approximation are treated as
additional noise. It is this non-integer penalty that ultimately limits
the performance of this scheme in the high signal-to-noise ratio (SNR)
regime. This obstacle was overcome in \cite{niesen12} in the high SNR
limit by combining compute-and-forward with the rational alignment
scheme due to Motahari \emph{et al.} \cite{mgmk09}. 

For the time-varying channels considered here, we propose a new scheme,
termed \emph{computation alignment}, that allows for a much sharper
analysis at finite SNRs. Our scheme combines compute-and-forward with a
signal-alignment scheme inspired by ergodic interference alignment
\cite{ngjv12IT}. By carefully matching channel realizations, our
approach decomposes the wireless channel with time-varying
complex-valued channel gains into subchannels with constant
integer-valued channel gains, over which lattice codes can be employed
efficiently.

\subsection{Related Work}
\label{sec:intro_related}

Relay networks have been the subject of considerable interest. For
\emph{wired} networks (i.e., networks of point-to-point channels),
Koetter \emph{et al.} recently proved that it is capacity-optimal to
separate channel and network coding \cite{kem11}. It is now well known
that routing over the resulting graph of bit pipes is optimal for
unicasting \cite{ff56,efs56} and, as demonstrated by Ahlswede \emph{et
al.} \cite{acly00}, network coding is required to achieve the multicast
capacity. 

For \emph{wireless} networks, channel-network separation is not always
optimal: higher rates can be achieved using more sophisticated relaying
techniques such as decode-and-forward (see, e.g.,
\cite{ce79,ltw04,kgg05}) compress-and-forward (see, e.g.,
\cite{ce79,kgg05,ssps09,avestimehr11,lkec11}), amplify-and-forward (see,
e.g., \cite{sg00,ltw04,gv05,bzg07,mgm10}), and compute-and-forward
(see, e.g., \cite{nazer11a,wnps10,ncl10,fsk11,niesen12}). While for
certain classes of deterministic networks the unicast and multicast
capacity regions are known \cite{arefphd,rk06,avestimehr11}, in the
general, noisy case, these problems remain open.  Recent progress has
been made by focusing on finding capacity approximations
\cite{etw08,avestimehr11,bpt10,mdgt11,niesen10b}. 

As mentioned above, our approach combines signal alignment with lattice
coding techniques. Signal alignment for interference management has
proved useful especially for the Gaussian interference channel
\cite{mmk08,cadambe08,bpt10,sjvjs08,ngjv12IT,mgmk09}. In particular,
ergodic alignment has been used to show that half the interference-free
rate is achievable at any SNR \cite{ngjv12IT} as well as derive sharper
scaling laws for ad-hoc networks \cite{niesen11}. More recently, several
groups have used alignment to make progress on the multiple unicast
problem in wireless networks \cite{jcj10,gjjc10,sa11,wgj11}. 

Lattice codes provide an elegant framework for many classical Gaussian
multi-terminal problems \cite{zse02,zamir09}. Beyond this role, it has
recently been shown that they have a central part to play in approaching
the capacity of networks that include some form of interference
\cite{nazer11a,pzek09,bpt10,wnps10,ncl10,ncl09,sjvjs08}.

\subsection{Organization}
\label{sec:intro_organization}

The remainder of this paper is organized as follows.
Section~\ref{sec:problem} introduces the problem setting as well as
notation. Section~\ref{sec:main} presents the main results as well as a
motivating example that captures the key features of the computation
alignment scheme.
Sections~\ref{sec:proofs_quantization}--\ref{sec:proofs_multiple}
provide detailed proofs for our main results.
Section~\ref{sec:conclusions} concludes the paper.

\section{Problem Setting and Notation}
\label{sec:problem}

This section formally introduces the problem setting and notation.
Although we are interested here in relay networks with several layers,
it will be convenient to first discuss networks with a single layer.
This single-layer network model is presented in
Section~\ref{sec:problem_single}. We then apply the insights obtained
for networks with a single layer of relays to networks with more than
one layer of relays. This multi-layer network model is presented in
Section~\ref{sec:problem_multiple}. Before we formally describe these
two problem settings, we introduce some notational conventions in
Section~\ref{sec:problem_notation}.

\subsection{Notational Conventions}
\label{sec:problem_notation}

Throughout this paper, $\log(\cdot)$ denotes the logarithm to the base
two, and all capacities and rates are hence expressed in terms of bits.
We use bold font lower and upper case, such as $\bm{h}$ and $\bm{H}$, to
denote vectors and matrices, respectively. Whenever the distinction is
of importance, realizations of random variables will be denoted by
sans-serif font, e.g., $\bm{\msf{H}}$ is a realization of the random
matrix variable $\bm{H}$.

\subsection{Single-Layer Relay Networks}
\label{sec:problem_single}

\begin{figure}[htbp]
    \centering
    \includegraphics{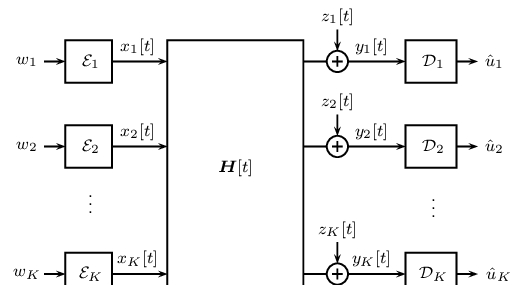}
    \caption{$K$ transmitters communicate an invertible set of functions
    $u_k = f_k(w_1,w_2,\ldots, w_K)$ of their messages to $K$ receivers
    over a time-varying interference channel.} 
    \label{fig:single}
\end{figure}

We start with a model for a wireless relay network with a single layer.
This single layer is to be interpreted as a part of a larger relay
network, to be introduced formally in
Section~\ref{sec:problem_multiple}. The single-layer relay network
consists of $K$ transmitters and $K$ receivers as depicted in
Fig.~\ref{fig:single}. We think of the $K$ transmitters as being located
at either the source nodes or at the relay nodes in some layer, say $d$,
of the larger relay network. We think of the $K$ receivers as being
located at the relay nodes at layer $d+1$ of the larger relay network. 

Each transmitter, indexed by $k \in \{1,\ldots, K\}$, has access to a
\emph{message} $w_k$ that is generated independently and uniformly over
$\{1,\ldots, 2^{TR_k}\}$, where $R_k$ is the \emph{rate} of transmitter
$k$. Each receiver, indexed by $m \in \{1,\ldots, K\}$, aims to recover
a \emph{deterministic function}
\begin{equation*}
    u_m \defeq f_m(w_1, \ldots, w_K)
\end{equation*}
of the $K$ messages $(w_1, \ldots, w_K)$.  We impose that the functions
$(f_m)_{m=1}^K$ computed at the receivers are invertible. In
other words, there must exist a function $g$ such that
$g(u_1,u_2,\ldots, u_K)=(w_1,w_2, \ldots, w_K)$.  Since the
functions to be computed at the receivers are deterministic, noise
is prevented from accumulating as messages traverse the larger relay
network.  Moreover, since the functions to be computed are invertible,
no information is lost from one layer to the next in the larger relay
network.

The transmitters communicate with the receivers over a Rayleigh-fading
complex Gaussian channel modeled as follows.  The \emph{channel output}
$y_m[t]\in\C$ at receiver $m\in\{1,\ldots, K\}$ and time $t\in\N$ is
given by 
\begin{equation}
    \label{eq:channel}
    y_m[t] \defeq \sum_{k=1}^K h_{m,k}[t]x_k[t] + z_m[t],
\end{equation}
where $x_k[t]\in\C$ is the \emph{channel input} at transmitter $k$,
$h_{m,k}[t]$ is the \emph{channel gain} between transmitter $k$ and
receiver $m$, and $z_m[t]$ is \emph{additive receiver noise}, all at time $t$.
The noise $z_m[t]$ is circularly-symmetric complex
Gaussian with mean zero and variance one, and independent of the
channel inputs $x_k[t]$ for $k\in\{1,\ldots, K\}$, $t\in\N$, and
independent of all other $z_{m^\prime}[t^\prime]$ for $(m^\prime,
t^\prime)\neq (m,t)$. Each channel gain $h_{m,k}[t]$ is assumed to be
circularly-symmetric complex Gaussian, with mean zero and variance one,
i.e., we assume Rayleigh fading. As a function of time $t$,
$(h_{m,k}[t])_{t\in\N}$ is a stationary ergodic process for every $m$
and $k$. The $K^2$ processes $(h_{m,k}[t])_{t\in\N}$ are mutually
independent as a function of $m,k$. Denoting by 
\begin{equation*}
    \bm{H}[t] \defeq (h_{m,k}[t])_{m,k}
\end{equation*}
the matrix of channel gains at time $t$, this implies that the matrix process
\begin{equation*}
    \bm{H}[1], \bm{H}[2], \bm{H}[3], \dots
\end{equation*}
is also stationary and ergodic. The channel gains $\bm{H}[t]$ are known
at all nodes in the network at time $t$. In other words, we assume
availability of full instantaneous channel-state information (CSI) throughout
the network. 

Each transmitter consists of an \emph{encoder} $\mc{E}_k$ mapping its
message $w_k$ into a sequence of $T$ channel inputs
\begin{equation*}
    (x_k[t])_{t=1}^T \defeq \mc{E}_k(w_k),
\end{equation*}
satisfying an \emph{average power constraint}
\begin{equation*}
    \frac{1}{T} \sum_{t=1}^T\abs{x_k[t]}^2 \leq P. 
\end{equation*}
Each receiver consists of a \emph{decoder} $\mc{D}_m$ mapping its
observed channel output into an estimate 
\begin{equation*}
    \hat{u}_m \defeq \mc{D}_m(y_m[1],\ldots,y_m[T])
\end{equation*}
of the desired function $u_m = f_m(w_1,\ldots,w_K)$. 
The average probability of error across all relays is defined as
\begin{align*}
    \Pp\big( {\textstyle\bigcup_{m=1}^K} \{ \hat{u}_m \neq u_m \} \big).
\end{align*} 

\begin{definition} 
    A computation sum rate $R(P)$ is \emph{achievable} if, for every
    $\varepsilon > 0$ and every large enough $T$, there exist encoders
    with blocklength $T$, average power constraint $P$, and rates
    satisfying $\sum_{k=1}^K R_k \geq R(P)$, and there exist decoders
    computing some invertible deterministic function $(f_m)_{m=1}^K$
    with average probability of error at most $\varepsilon$.  The
    \emph{computation sum capacity} $C(P)$ of the single-layer relay
    network is the supremum of all achievable computation sum rates
    $R(P)$.
\end{definition}

Observe that the definition of computation sum capacity does not
prescribe the function of the messages to be computed at the receivers.
The only requirement is that these functions are deterministic and
invertible.  In other words, the computation sum capacity is the largest
sum rate at which \emph{some} (as opposed to a \emph{specific}) function
can be reliably computed.

\subsection{Multi-Layer Relay Networks}
\label{sec:problem_multiple}

\begin{figure*}[htbp]
    \centering
    \includegraphics{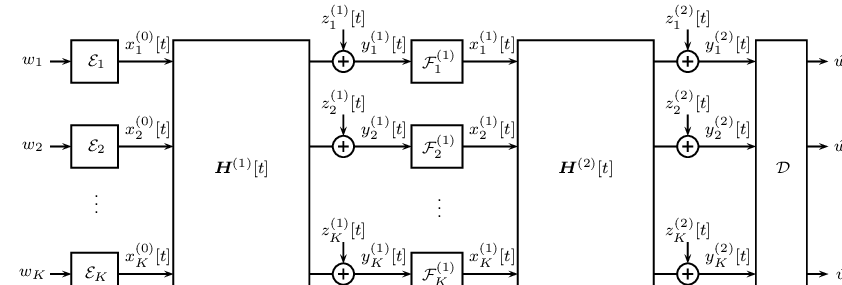}
    \caption{A multi-layer relay network with $D=2$ layers and $K$ relays
    per layer.}  
    \label{fig:multiple}
\end{figure*}

Having described the single-layer network setting, we now turn to
networks with multiple layers of relays. These networks consist of a
concatenation of $D$ single-layer networks as defined in
Section~\ref{sec:problem_single}. The network contains $K$ source nodes
at layer zero connected through a Rayleigh-fading channel to $K$ relay
nodes at layer one. Layer $d$ in the network contains $K$ relay nodes
connected through a Rayleigh-fading channel to $K$ relay nodes at layer
$d+1$. The relay nodes at layer $D$ are connected to the destination
node at layer $D+1$ through orthogonal bit pipes of infinite capacity.
This ensures that the intermediate relay layers, not the bit pipes, are
the bottleneck in the network (see also the remark below).  This
scenario is depicted in Fig.~\ref{fig:multiple}.

Formally, each transmitter at a source node, indexed by $k\in\{1, 
\ldots, K\}$, has access to a \emph{message} $w_k$ of rate $R_k$ that is
generated independently and uniformly over $\{1, \ldots, 2^{TR_k}\}$.
The receiver at the destination node aims to recover the transmitted
messages $(w_1, w_2, \ldots, w_K)$. 

The transmitters at layer $d\in\{0, \ldots, D-1\}$ communicate with
the receivers at layer $d+1$ over a Rayleigh-fading complex Gaussian
channel modeled as in the single-layer case. The \emph{channel output} 
$y_m^{(d+1)}[t]\in\C$ at the receiver at relay $m\in\{1,\ldots, K\}$ in
layer $d+1$ and time $t\in\N$ is given by 
\begin{equation*}
    y_m^{(d+1)}[t] 
    \defeq \sum_{k=1}^K h_{m,k}^{(d+1)}[t]x_k^{(d)}[t] + z_m^{(d+1)}[t],
\end{equation*}
where $x_k^{(d)}[t]$ is the \emph{channel input} at the transmitter at
relay or source $k\in\{1,\ldots, K\}$ at layer $d$. The \emph{channel
gains} $h_{m,k}^{(d+1)}[t]$ and the \emph{additive noise} $z_m^{(d+1)}$
satisfy the same statistical assumptions as in the single-layer network
described by~\eqref{eq:channel}, and they are assumed to be independent
across different layers. 

As mentioned earlier, the relay nodes in layer $D$ are connected to the
destination node at layer $D+1$ through $K$ orthogonal bit pipes with
infinite capacity. Without loss of generality, we can assume that the
relays in layer $D$ simply forward their observed channel outputs
$y_m^{(D)}[t]$ to the destination node.

\begin{remark}
    \label{rem:pipes}
    The bit pipes from the final relay layer to the destination can be
    replaced with another (symmetric) multiple-access channel model
    without affecting our main results. We have used a model with
    orthogonal links with infinite capacity in order to focus on the
    case when the capacity bottleneck occurs between relay layers, not
    in the final hop. 
\end{remark}

Each transmitter at source node $k$ consists of an \emph{encoder}
$\mc{E}_k$ mapping its message $w_k$ into a sequence of $T$ channel inputs, 
\begin{equation*}
    (x_k^{(0)}[t])_{t=1}^T \defeq \mc{E}_k(w_k),
\end{equation*}
satisfying an average power constraint of $P$.

The receiver-transmitter pair at relay node $k$ in layer $d\in\{1,\ldots,
D-1\}$ consists of a \emph{relaying function} $\mc{F}_k^{(d)}$ mapping
the block of observed channel outputs $(y_m^{(d)}[1], \ldots,
y_m^{(d)}[T])$ from layer $d$ into a block of channel 
inputs 
\begin{equation*}
    (x_k^{(d)}[t])_{t=1}^T 
    \defeq \mc{F}_k^{(d)}( y_m^{(d)}[1], \ldots, y_m^{(d)}[T])
\end{equation*}
for layer $d+1$, satisfying an average power constraint of
$P$.\footnote{As may be seen from the definition of the relaying
function, we do not impose causality for the operations at the relay.
This assumption is only for ease of notation---since we are dealing with
a layered network, all results are also valid for causal relaying
functions by coding over several blocks.}

Finally, the receiver at the destination node in layer $D+1$ consists of
a \emph{decoder} $\mc{D}$ mapping its observed channel outputs
(forwarded from the relays at layer $D$) into an estimate 
\begin{equation*}
    (\hat{w}_1, \hat{w}_2, \ldots, \hat{w}_K) 
    \defeq \mc{D}\big((y_1^{(D)}[1],\ldots,y_1^{(D)}[T]), 
    \ldots, (y_K^{(D)}[1],\ldots,y_K^{(D)}[T])\big)
\end{equation*}
of the messages $(w_1, \ldots, w_K)$. The average probability of
error is defined as
\begin{align*}
    \Pp\big( {\textstyle\bigcup_{k=1}^K} \{ \hat{w}_k \neq w_k \} \big).
\end{align*} 

\begin{definition}
    A sum rate $R^{(D)}(P)$ is \emph{achievable} if, for every
    $\varepsilon > 0$ and every large enough $T$, there exist encoders,
    relaying functions, and a decoder with blocklength $T$, average
    power constraint $P$, rates satisfying $\sum_{k=1}^K R_k \geq
    R^{(D)}(P)$, and average probability of error at most $\varepsilon$.
    The \emph{sum capacity} $C^{(D)}(P)$ of the multi-layer relay
    network is the supremum of all achievable sum rates $R^{(D)}(P)$.
\end{definition}

\section{Main Results}
\label{sec:main}

We now state our two main results, an approximate characterization of
the computation sum capacity $C(P)$ of the single-layer relay network
(Section~\ref{sec:main_single}) and an approximate characterization of
the sum capacity $C^{(D)}(P)$ of the $D$-layer relay network
(Section~\ref{sec:main_multiple}), both under i.i.d. Rayleigh fading.
The proofs will be presented in detail in
Sections~\ref{sec:proofs_quantization}--\ref{sec:proofs_multiple}. In
Section~\ref{sec:motivate}, we explore a simple example that captures
the intuition behind our computation-alignment scheme used to prove the
main results.

\subsection{Single-Layer Relay Networks}
\label{sec:main_single}

We start with the analysis of the computation sum capacity $C(P)$ of a
single-layer relay network consisting of $K$ source nodes and $K$ relay
nodes. 

\begin{theorem}
    \label{thm:compute} 
    For a single-layer network with $K$ source nodes, $K$ relay nodes,
    and time-varying i.i.d. Rayleigh channel coefficients, the
    computation sum capacity $C(P)$ is lower and upper bounded as 
    \begin{equation*}
        K\log(P)-7K^3
        \leq C(P) 
        \leq K\log(P)+5K\log(K)
    \end{equation*} 
    for every power constraint $P \geq 1$.
\end{theorem}

The proof of the lower bound in Theorem~\ref{thm:compute} is presented
in Sections~\ref{sec:proofs_K2} (for $K=2$) and \ref{sec:proofs_K3} (for
$K > 2$). The proof of the upper bound in Theorem~\ref{thm:compute} is
presented in Section~\ref{sec:proofs_upper}.

Theorem~\ref{thm:compute} provides an approximate characterization of
the computation sum capacity $C(P)$ of the single-layer relay network.
Comparing the upper and lower bounds shows that the approximation is up
to an additive gap of $7K^3+5K\log(K)$ bits/s/Hz. In particular, the
gap does not depend on the power constraint $P$. In other words,
Theorem~\ref{thm:compute} asserts that 
\begin{equation*}
    C(P) = K\log(P) \pm O(1).
\end{equation*}
This is considerably stronger than the best previously
known bounds in \cite{niesen12} on the computation sum capacity of such
networks, which only provide the degrees-of-freedom approximation
\begin{equation*}
    C(P) = K\log(P) \pm o(\log(P))
\end{equation*}
as $P\to\infty$.

The upper bound in Theorem~\ref{thm:compute} results from the cut-set
bound, allowing cooperation among the sources and among the relays. This
transforms the channel into a $K\times K$ multiple-input multiple-output
system, and the upper bound follows from analyzing its capacity. From
Theorem~\ref{thm:compute}, we hence see that computation of a (carefully
chosen) invertible function can be performed in a distributed manner
with at most a $O(1)$ loss in rate compared to the centralized scheme in
which the $K$ transmitters cooperate and the $K$ receivers cooperate. 

The communication scheme achieving the lower bound in
Theorem~\ref{thm:compute} is based on a combination of a lattice
computation code with a signal-alignment strategy, which we term
\emph{computation alignment}. We now provide a brief description of
these two components and how they interact---the details of the argument
can be found in the proof of Theorem~\ref{thm:compute} in
Sections~\ref{sec:proofs_K2} and \ref{sec:proofs_K3}.

A \emph{lattice} is a discrete subgroup of $\R^T$, and hence has the
property that any integer combination of lattice points is again a
lattice point. A \emph{lattice computation code} as defined in
\cite{nazer11a} uses such a lattice, intersected with an appropriate
bounding region to satisfy the power constraint, as its codebook. This
strategy is designed for the case where the channel coefficients remain
constant over the duration of the codeword, $h_{m,k}[t] = h_{m,k}$.
Assume for the moment that the channel gains are all integers. Then each
receiver observe an integer combination of codewords plus Gaussian
noise. By the lattice property, this is equal to some other codeword
plus noise. If the lattice is carefully chosen, the receivers can remove
the noise, and are hence left with the integer combination of the
codewords which corresponds to a deterministic function of the messages. 

In general, the channel coefficients will not be integer multiples of
one another. In this case, each receiver may aim to decode an integer
combination of codewords that best approximate the linear combination
produced by the channel. \cite[Theorem~3]{nazer11a} states that the
receivers can decode integer combinations with coefficients $a_{m,k} \in
\Z + \sqrt{-1}\Z$ if the rates (from the transmitters) satisfy
\begin{align}
    \label{eq:latticerate}
    R_k 
    < \min_{k: a_{m,k} \neq 0} \max_{\alpha_m \in \C} 
    \log\biggl(\frac{P}{\alpha_m^2 + P \sum_k \abs{\alpha_m h_{m,k} -
    a_{m,k}}^2 }\biggr). 
\end{align} 

From the denominator in~\eqref{eq:latticerate}, we see that the
performance of this lattice-coding approach is closely tied to how well
the channel gains $h_{m,k}$ can be approximated by integers. If
$h_{m,k}$ is not a rational, then this approximation cannot be done
perfectly, resulting in significant rate loss especially for larger
values of power $P$ as shown in \cite{niesen12}. Using lattices by
itself as described above is hence not sufficient to prove a
constant-gap result as in Theorem~\ref{thm:compute}.

Instead, in this paper we combine lattice codes with an alignment scheme
inspired by ergodic interference alignment \cite{ngjv12IT}. By
exploiting the time-varying nature of the channels, we code over several
channel uses to create subchannels with integer coefficients over which
lattice codes can then be efficiently used. We term this combination of
alignment and lattice codes \emph{computation alignment}. Below, we
discuss a simple example of our scheme that elucidates some of the key
features of the general construction.

\subsection{Motivating Example} 
\label{sec:motivate}

The computation-alignment scheme is best illustrated for $K=2$ users.
Consider a time slot $t_1$ and consider the four channel gains
$h_{m,k}[t_1]$ at time $t_1$. For simplicity (and without too much loss
of generality), assume that
\begin{align*}
    h_{1,1}[t_1] & =
    h_{1,2}[t_1] =
    h_{2,1}[t_1] = 1, \\
    h_{2,2}[t_1] & = h
\end{align*}
for some $h\in\C$. If we communicate over only time slot $t_1$ alone,
the channel outputs are
\begin{align*}
    y_1[t_1] & = x_1[t_1]+x_2[t_2]+z_1[t_1], \\
    y_2[t_1] & = x_1[t_1]+hx_2[t_2]+z_2[t_1].
\end{align*}
Since the channel gains to receiver one are both integers, lattice codes
can be used to efficiently compute a linear combination of the
transmitted codewords. On the other hand, for most values of $h$,
lattice codes as described above can not be used for efficient
computation at receiver two.  As a result, over one time slot, we can
only reliably compute invertible functions of \emph{one} data stream.
This yields a computation sum rate of roughly $\log(P)$.

We now argue that if we code over $t_1$ and a second, carefully matched,
time slot $t_2$, we can in fact reliably compute invertible functions of
\emph{three} data streams. This yields a computation sum rate of roughly
$\tfrac{3}{2}\log(P)$. Assume we can find a second time slot $t_2$ such
that\footnote{While we consider only a single pair $(t_1,t_2)$ of time
slots, it can be shown that with high probability almost all time slots
can be matched such that these conditions are (approximately)
satisfied.}
\begin{align*}
    h_{1,1}[t_2] & =
    h_{1,2}[t_2] = 1, \\
    h_{2,1}[t_2] & = -1, \\
    h_{2,2}[t_2] & = h.
\end{align*}
Over the two time slots, $t_1$ and $t_2$, the channel outputs are
\begin{align*}
    \bm{y}_1 & \defeq 
    \begin{pmatrix}
        y_1[t_1] \\
        y_1[t_2]
    \end{pmatrix}
    = 
    \begin{pmatrix}
        x_1[t_1] \\
        x_1[t_2]
    \end{pmatrix}
    + 
    \begin{pmatrix}
        x_2[t_1] \\
        x_2[t_2]
    \end{pmatrix}
    +
    \begin{pmatrix}
        z_1[t_1] \\
        z_1[t_2]
    \end{pmatrix}, \\
    \bm{y}_2 & \defeq
    \begin{pmatrix}
        y_1[t_1] \\
        y_1[t_2]
    \end{pmatrix}
    = 
    \begin{pmatrix}
        x_1[t_1] \\
        -x_1[t_2]
    \end{pmatrix}
    + h
    \begin{pmatrix}
        x_2[t_1] \\
        x_2[t_2]
    \end{pmatrix}
    +
    \begin{pmatrix}
        z_2[t_1] \\
        z_2[t_2]
    \end{pmatrix}.
\end{align*}

Over this block channel, transmitter one aims to send symbols
$s_{1,1}$ and $s_{1,2}$ and transmitter two aims to send symbol
$s_{2,1}$. These symbols are mapped onto the two time slots using
transmit vectors $\bm{v}_{1,1}$, $\bm{v}_{1,2}$, and $\bm{v}_{2,1}$,
i.e., 
\begin{align*}
    \begin{pmatrix}
        x_1[t_1] \\
        x_1[t_2]
    \end{pmatrix}
    & = \bm{v}_{1,1} s_{1,1} + \bm{v}_{1,2} s_{1,2} \\
    \begin{pmatrix}
        x_2[t_1] \\
        x_2[t_2]
    \end{pmatrix}
    & = \bm{v}_{2,1} s_{2,1}. 
\end{align*}
We now describe how to choose these transmit vectors.

\begin{figure}[htbp]
    \centering
    \hspace{-2.05cm}\includegraphics{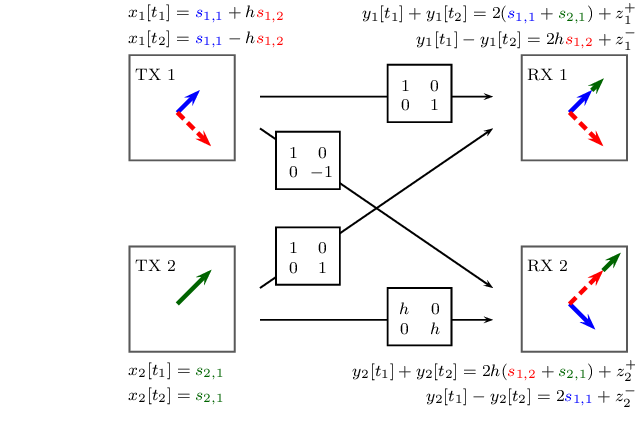}
    \caption{Computation alignment scheme for two users over two slots.
    Transmitter $1$ sends symbols $s_{1,1}$ and $s_{1,2}$ from two
    independent lattice codewords while transmitter $s_{2,1}$ sends one
    symbol from a single lattice codeword. After appropriate scaling,
    receiver observes the sum of two symbols in one subchannel and the
    remaining symbol in the other subchannel. Put together, these integer
    combinations form a full rank set of linear equations. In the figure, 
    $z_{k}^+ \defeq z_{k}[t_1] + z_{k}[t_2]$ and $z_k^- \defeq
    z_{k}[t_1] - z_{k}[t_2]$. }  
    \label{fig:motivate}
\end{figure}

We begin with the special case where $\abs{h}=1$.  We choose the
transmit vectors to be $\bm{v}_{1,1} = (1~~1)^{\T}$, $\bm{v}_{1,2} = h(1
~ -1)^{\T}$, and $\bm{v}_{2,1} = (1~~1)^{\T}$.  This leads to the
effective channel 
\begin{align*} 
    \bm{y}_1
    = 
    \begin{pmatrix}
        1\\
        1
    \end{pmatrix}
    ( s_{1,1} + s_{2,1} ) +  h
    \begin{pmatrix}
        1\\
        -1
    \end{pmatrix}
    s_{1,2} + \bm{z}_1, \\
    \bm{y}_2 = h
    \begin{pmatrix}
        1\\
        1
    \end{pmatrix}
    ( s_{1,2} + s_{2,1} ) +  
    \begin{pmatrix}
        1\\
        -1
    \end{pmatrix}
    s_{1,1} + \bm{z}_2. 
\end{align*} 
Thus, each receiver sees two orthogonal subchannels, each carrying
\emph{integer} combinations of symbols. Receiver one observes the sum
$s_{1,1} + s_{2,1}$ on one subchannel and $s_{1,2}$ on the other;
receiver two observes the sum $s_{1,2}+s_{2,1}$ on one subchannel and
$s_{1,1}$ on the other. We say that the subchannels are \emph{aligned}
for efficient computation in that they are orthogonal and have integer
coefficients. Given the orthogonality of the subchannels, they can be
recovered at both receivers using matched filters. And given that all
subchannels have integer coefficients, lattice codes can be efficiently
employed to achieve a computation sum rate of roughly
$\frac{3}{2}\log(P)$. See Fig.~\ref{fig:motivate} for an illustration.

Next, consider the case $\abs{h} < 1$ (the case $\abs{h} > 1$ can be
dealt with similarly). In this setting, one can improve upon the scheme
above by steering the effective channel gains of aligned symbols to the
nearest integer, rather than fully equalizing them. Let $b$ be the
smallest natural number such that 
\begin{align*}
    1 \leq b\abs{h} < 2,
\end{align*} 
and set the transmit vectors to be $\bm{v}_{1,1} = (1~~1)^{\T}$,
$\bm{v}_{1,2} = bh(1 ~ -1)^{\T}$, and $\bm{v}_{2,1} = (1~~1)^{\T}$.  The
key observation here is that, since $b\abs{h}\in[1,2)$, all transmit vectors
have comparable lengths, leading to a better power allocation across
subchannels than the same choice of transmit vectors with $b=1$.

With this, the effective channel becomes
\begin{align*}
    \bm{y}_1 & = 
    \begin{pmatrix}
        1\\
        1
    \end{pmatrix}
    ( s_{1,1} + s_{2,1} ) 
    + h 
    \begin{pmatrix}
        1\\
        -1
    \end{pmatrix}
    b s_{1,2} + \bm{z}_1, \\
    \bm{y}_2 & = h
    \begin{pmatrix}
        1\\
        1
    \end{pmatrix}
    ( b s_{1,2} + s_{2,1} ) +  
    \begin{pmatrix}
        1\\
        -1
    \end{pmatrix}
    s_{1,1} + \bm{z}_2. 
\end{align*}
Since $b$ is an integer, this is again aligned for efficient
computation and achieves the same computation sum rate of roughly
$\tfrac{3}{2}\log(P)$. 

Building on this example, the general scheme developed in
Section~\ref{sec:proofs_K2} encodes $2L -1$ data streams across $L$ time
slots to reach a computation sum rate of approximately
$\tfrac{2L-1}{L}\log(P)$. By taking $L \rightarrow \infty$, this
strategy can approach the desired computation sum rate $2\log(P)$ to
within a constant gap. As shown in Section~\ref{sec:proofs_K3}, we can
establish aligned subchannels for $K > 2$ users in a similar fashion.

\subsection{Multi-Layer Relay Networks}
\label{sec:main_multiple}

Having analyzed the computation sum capacity for single-layer relay
networks, we now turn to the sum capacity of relay networks with
multiple layers. Unlike the single-layer network, there is only one
destination node, which is interested in recovering the original
messages (and not merely a function of them). We are hence interested
here in sum capacity in the traditional sense. 

\begin{theorem}
    \label{thm:multiple}
   Consider a multi-layer relay network with $D\geq 1$ layers, $K\geq 2$
   source nodes, and $K$ relay nodes per layer. If the channel 
   coefficients are time-varying and i.i.d. Rayleigh, the sum capacity
   $C^{(D)}(P)$ is lower and upper bounded as
    \begin{equation*}
        K\log(P)-7K^3
        \leq C^{(D)}(P) 
        \leq K\log(P)+5K\log(K)
    \end{equation*}
    for every power constraint $P \geq 1$. 
\end{theorem}

The proof of Theorem~\ref{thm:multiple} is presented in
Section~\ref{sec:proofs_multiple}. The upper bound follows directly from
the same cut-set bound argument as in Theorem~\ref{thm:compute}. The
lower bound uses compute-and-forward in each layer as analyzed in
Theorem~\ref{thm:compute}. The destination node gathers all the computed
functions and inverts them to recover the original messages sent by the
source nodes. 

Theorem~\ref{thm:multiple} provides an approximate characterization of
the sum capacity of the $D$-layer relay network. The gap between the
lower and upper bounds is $7K^3+5K\log(K)$ bits/s/Hz as in
Theorem~\ref{thm:compute}. This gap is again independent of the power
constraint $P$, showing that
\begin{equation*}
    C^{(D)}(P) = K\log(P)\pm O(1).
\end{equation*}
Moreover, the gap in Theorem~\ref{thm:multiple} is also independent of
the network depth $D$. In other words, the approximation guarantee is
uniform in the network parameter $D$. 

It is interesting to compare this approximation result to other known
capacity approximations for general Gaussian relay networks of the form
considered here. For general relay networks, these bounds rely on a
compress-and-forward scheme and achieve an additive approximation gap
of $1.26(D+1)K$ bits/s/Hz \cite{avestimehr11,lkec11}. Unlike the gap in
Theorem~\ref{thm:multiple}, this gap is not uniform in the network depth
$D$. This is due to the use of compress-and-forward: In each relay
layer, the channel output, consisting of useful signal as well as
additive noise, is quantized and forwarded to the next layer. Thus, with
each layer additional noise accumulates, degrading performance as the
network depth increases. The result is an approximation guarantee that
becomes worse with increasing network depth. 

Theorem~\ref{thm:multiple} in this paper avoids this difficulty by
completely removing channel noise at each layer in the network.  This is
achieved by decoding a deterministic (and hence noiseless) function of
the messages at each relay. Thus, noise is prevented from accumulating
as the messages traverse the network. It is this feature of
compute-and-forward that enables the uniform approximation guarantee in
Theorem~\ref{thm:multiple}.

We remark that the $7K^3$ term in the lower bound of
Theorem~\ref{thm:multiple} is due to the construction ensuring that all
received signals are integer multiples of each other. If instead of
Rayleigh fading we consider channel gains with equal magnitude and
independent uniform phase fading, the lower bound in
Theorem~\ref{thm:multiple} can be sharpened to $K\log(P)$, resulting in
an approximation gap of $5K\log(K)$. Deriving
capacity approximations with better dependence on $K$ for general fading
processes is an interesting direction for future work.

It is also worth mentioning that, unlike the gap presented here, the
approximation gap in \cite{avestimehr11} is uniform in the fading
statistics. Developing communication schemes that guarantee an
approximation gap that is uniform in both the network depth and the
fading statistics is therefore of interest. 

Finally, like other signal alignment schemes for time-varying channels
such as \cite{cadambe08} and \cite{ngjv12IT}, the communication scheme
proposed in this paper suffers from long delays. This limits the
practicality of these schemes even for moderate values of $K$. Finding
ways to achieve signal alignment (be it for interference management or
function computation) with less delay is hence of importance.

\section{Channel Quantization} 
\label{sec:proofs_quantization}

The achievable scheme in Theorem~\ref{thm:compute} groups together time
slots so that an appropriate linear combination of the channel outputs
within each group yields a more desirable effective channel. This
grouping of time slots is performed such that the corresponding channel
realizations ``match'' in a sense to be made precise later. Since each
possible channel realization has measure zero, we cannot hope for
channel matrices to match exactly. Instead, we will look for channel
matrices that approximately match. This approximate matching is
described by considering a quantized version of the channel gains. In
this section, we describe such a quantization scheme, similar to the one
used for ergodic interference alignment in \cite{ngjv12IT}.

\begin{figure}[htbp]
    \centering
    \scalebox{0.615}{\includegraphics{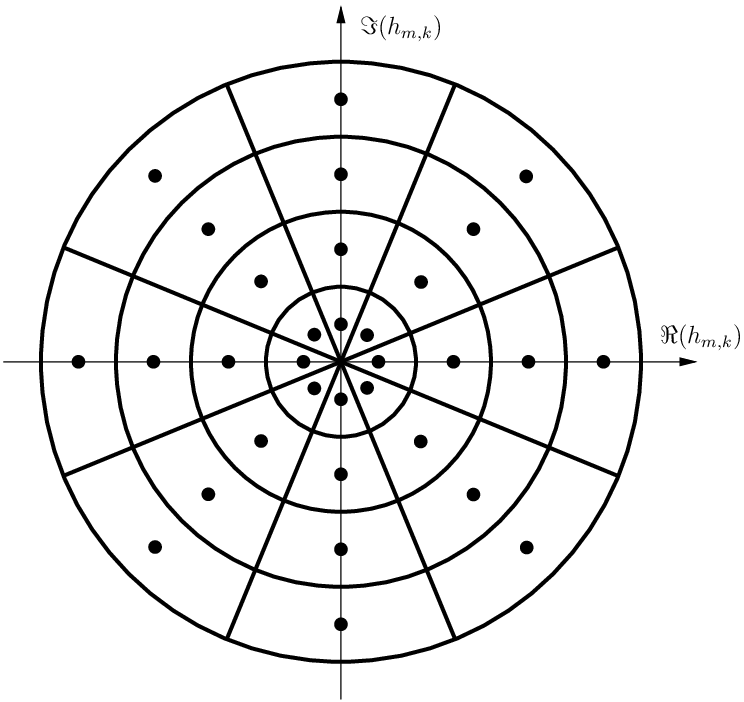}} 
    \caption{Quantization scheme for channel coefficients. Coefficients
    up to magnitude $\nu$ are quantized by magnitude and angle.  The
    number of angular regions is a multiple of $L$ to ensure that
    multiplying a quantization point by any $L$\/th root of unity
    results again in a quantization point. In the figure, $\nu=2$
    and $L=2$} 
    \label{fig:quant}
\end{figure}

We divide the complex plane from the origin up to distance $\nu$ into
concentric rings centered at the origin and with spacing  $1/\nu$ for
some natural number $\nu \geq 2$ to be chosen later. Then, we divide
each of these $\nu^2$ rings into $\nu^2 L$ segments with identical
central angles of size $2\pi/(\nu^2 L)$ for some $L \in \N$ also to be
chosen later. These segments serve as quantization cells for the channel
coefficients. Each segment is represented by the mid-point on the
bisector of the corresponding central angle (see Fig. \ref{fig:quant}).
We add one additional quantization point at infinity to which we will
map all channel gains with magnitude larger than $\nu$. Note that
multiplying a quantization point by any $L$\/th root of unity results
again in a quantization point. We will use this property frequently in
the sequel.

Let $\hat{h}_{m,k}[t]$ denote the quantized version of the channel
coefficient $h_{m,k}[t] \in \C$. We then have that $\hat{h}_{m,k}[t] =
\infty$ if $\abs{h_{m,k}[t]} > \nu$, and that $\hat{h}_{m,k}[t]$ is the
point in the ``middle'' of the quantization cell containing $h_{m,k}[t]$
otherwise (with ties broken arbitrarily).  We denote by $\hat{\mc{H}}$
the collection of all possible quantized channel values.  It will be
convenient in the following to denote by 
\begin{equation*}
    p_{\hat{\bm{H}}}(\hat{\bm{\msf{H}}})
    \defeq \Pp\big( \hat{\bm{H}}[1] = \hat{\bm{\msf{H}}}\big)
\end{equation*}
the probability mass function of the quantized channel gains
\begin{equation*}
    \hat{\bm{H}}[t] \defeq (\hat{h}_{m,k}[t])_{m,k}.
\end{equation*}

Note that the number of quantization regions is 
\begin{equation}
    \label{eq:hcard}
    \card{\hat{\mc{H}}} = \nu^4 L+1.
\end{equation}
By choosing $\nu$ large enough, we can ensure that the distance between
any point with magnitude less than $\nu$ and its closest quantization
point is arbitrarily small. In fact, for any $h_{m,k}[t]$ with
$\abs{h_{m,k}[t]} \leq \nu$, 
\begin{equation}
    \label{eq:quant}
    \abs{h_{m,k}[t]-\hat{h}_{m,k}[t]} \leq (\pi+1)/\nu.
\end{equation}
Furthermore, for any $\delta >0$, 
\begin{equation*}
    \Pp\big(\abs{h_{m,k}[t]} \leq \nu \ \forall m,k \big) \geq 1-\delta
\end{equation*}
for large enough $\nu$, and hence
\begin{equation}
    \label{eq:infty}
    \Pp\big(\abs{\hat{h}_{m,k}[t]} < \infty \ \forall m,k \big) \geq 1-\delta.
\end{equation}
Therefore \eqref{eq:quant} holds with probability at least $1-\delta$
for $\nu$ large enough. Finally, for any $h_{m,k}[t]$ such that
$\abs{h_{m,k}[t]}\leq \nu$,
\begin{equation}
    \label{eq:quant2}
    \max\big\{ \abs{\hat{h}_{m,k}[t]}, \abs{\hat{h}_{m,k}[t]}^{-1} \big\} 
    \leq 2 \max\big\{ \abs{h_{m,k}[t]}, \abs{h_{m,k}[t]}^{-1} \big\},
\end{equation}
since each finite quantization point is the mid-point of the
corresponding bisector interval.

Since the matrix process
\begin{equation*}
    \bm{H}[1], \bm{H}[2], \bm{H}[3], \ldots
\end{equation*}
is stationary and ergodic, the quantized process
\begin{equation*}
    \hat{\bm{H}}[1], \hat{\bm{H}}[2], \hat{\bm{H}}[3], \ldots
\end{equation*}
is also stationary and ergodic (see, e.g., \cite[Theorem~6.1.1,
Theorem~6.1.3]{durret04}). Moreover, since each $h_{m,k}[t]$ is
circularly symmetric, and since the quantization procedure preserves
this circular symmetry, the distribution of the quantized channel values
$\hat{h}_{m,k}[t]$ is invariant under multiplication by the $L$\/th root
of unity. Furthermore, since the $K^2$ processes $(h_{m,k}[t])_{t\in\N}$
are mutually independent as a function of $m,k$, so are the $K^2$
quantized processes $(\hat{h}_{m,k}[t])_{t\in\N}$. For future reference,
we summarize these observations in the following lemma.

\begin{lemma}
    \label{thm:symmetry}
    For each $m,k$, and $t$, the quantized channel gain $\hat{h}_{m,k}[t]$ 
    and its rotation $\exp\big(\sqrt{-1}\tfrac{2\pi}{L}\big)\hat{h}_{m,k}[t]$ 
    have the same distribution. The $K^2$ quantized processes 
    \begin{equation*}
        \hat{h}_{m,k}[1], \hat{h}_{m,k}[2], \hat{h}_{m,k}[3], \dots
    \end{equation*}
    are independent as a function of $m,k$. The quantized matrix 
    process
    \begin{equation*}
        \hat{\bm{H}}[1], \hat{\bm{H}}[2], \hat{\bm{H}}[3], \ldots
    \end{equation*}
    is stationary and ergodic. 
\end{lemma}

The basic idea behind our scheme is to match $L$ carefully chosen
time slots to create effective integer-valued channels. The most
intuitive version of this strategy is to match channels in a
``greedy'' fashion. However, it is simpler to analyze this strategy if
we split the block of $T$ time slots into $L$ consecutive subblocks and
assume that the $\ell$th time slot within a matched set always comes
from the $\ell$th subblock. This in turn allows us to draw upon the the
ergodic theorem to guarantee that each subblock contains roughly the
same number of each possible channel realization, meaning that almost
all channel realizations can be successfully matched.  Specifically,
consider a block of length $T$ of channel gains with $T$ a multiple of
$L$, and divide this block into $L$ subblocks each of length $T/L$.
Count the number of occurrences of a particular channel realization
$\hat{\bm{\msf{H}}}\in\hat{\mc{H}}^{K\times K}$ in one of the $L$
subblocks. By the ergodicity of the quantized matrix process, we expect
this number to be close to $T/L$ times the probability of this
realization. The next lemma formalizes this statement.

\begin{lemma}
    \label{thm:typical}
    For any $L,\nu\in \N$ and $\eta, \varepsilon > 0$, there exists
    $T=T(L,\nu)\in\N$ divisible by $L$ such that, with probability at
    least $1-\varepsilon$, we have, for all $\ell \in\{1, \ldots, L\}$,
    and all $\hat{\bm{\msf{H}}}\in\hat{\mc{H}}^{K\times K}$,
    \begin{equation*}
        \sum_{t=(\ell-1)T/L+1}^{\ell T/L}
        \ind\{ \hat{\bm{H}}[t] = \hat{\bm{\msf{H}}} \}
        \geq (1-\eta) p_{\hat{\bm{H}}}(\hat{\bm{\msf{H}}})T/L.
    \end{equation*}
\end{lemma}
\begin{IEEEproof}
    By Lemma~\ref{thm:symmetry}, the quantized matrix process
    \begin{equation*}
        \hat{\bm{H}}[1], \hat{\bm{H}}[2], \hat{\bm{H}}[3], \ldots
    \end{equation*}
    is stationary and ergodic. This stochastic process takes values in
    the finite set $\hat{\mc{H}}^{K\times K}$, and hence, by the ergodic
    theorem (see, e.g., \cite[Theorem~6.2.1]{durret04}), its empirical
    distribution converges to the true distribution almost surely. For
    fixed $\ell\in\{1,\ldots, L\}$, this implies that there exists a $T$
    such that with probability at least $1-\varepsilon/L$, we have for
    all $\hat{\bm{\msf{H}}}\in\hat{\mc{H}}^{K\times K}$,
    \begin{equation*}
        \sum_{t=(\ell-1)T/L+1}^{\ell T/L}
        \ind\{ \hat{\bm{H}}[t] = \hat{\bm{\msf{H}}} \}
        \geq (1-\eta)
        p_{\hat{\bm{H}}}(\hat{\bm{\msf{H}}} )T/L.
    \end{equation*}
    Applying the union bound over $\ell\in\{1,\ldots, L\}$ proves the
    result.
\end{IEEEproof}

\section{Proof of Lower Bound in Theorem~\ref{thm:compute} for Two Users}
\label{sec:proofs_K2}

In this section, we prove the lower bound in Theorem~\ref{thm:compute}
for the two-user case, i.e., $K=2$.  Consider a block of $T$ channel
gains, and divide this block into $L$ subblocks each of length of $T/L$
(which is assumed to be an integer).  The construction of the achievable
scheme in Theorem~\ref{thm:compute} consists of three main steps. First,
we carefully match $L$ time slots, one from each of the $L$ subblocks.
This matching is performed approximately $T/L$ many times such that
essentially all time slots in the block of length $T$ are matched (see
Section~\ref{sec:proofs_K2_matching}).  Second, we argue that any $L$
time slots matched in this fashion, when considered jointly, can be
transformed into parallel channels with (nearly) integer channel gains
using appropriate linear precoders at the transmitters and matched
filters at the receivers (see Section~\ref{sec:proofs_K2_precoding}).
Third, we show that over these integer channels we can efficiently and
reliably compute functions of the messages (see
Section~\ref{sec:proofs_K2_computation}).

\subsection{Matching of Channel Gains}
\label{sec:proofs_K2_matching}

We start with the matching step. Since the number of possible channel
realizations is uncountable, only approximate matching is possible. To
this end, we quantize each of the channel gains as described in
Section~\ref{sec:proofs_quantization}. Denote by $\hat{h}_{m,k}[t]$ the
quantized version of the channel gain $h_{m,k}[t]$. By
Lemma~\ref{thm:typical}, for every $\varepsilon_1>0$ and $\eta >0$,
there exists $T$ large enough such that with probability
$1-\varepsilon_1$, each of the $L$ subblocks is ``typical'', in the
sense that, for every subblock $\ell \in\{1, \ldots, L\}$, and every
realization $\hat{\bm{\msf{H}}}\in\hat{\mc{H}}^{K\times K}$ of
the quantized channel gains,
\begin{equation*}
    \sum_{t=(\ell-1)T/L+1}^{\ell T/L}
    \ind\{ \hat{\bm{H}}[t] = \hat{\bm{\msf{H}}} \}
    \geq (1-\eta) p_{\hat{\bm{H}}}(\hat{\bm{\msf{H}}})T/L.
\end{equation*}

Recall that full CSI is available at all transmitters and receivers.
Hence all transmitters and receivers can determine at the end of the
block of length $T$ if the realization of quantized channel gains is
typical. Whenever this is not the case, the decoders declare an error.
By the argument in the last paragraph, this happens with probability at
most $\varepsilon_1$. We assume in the following discussion that the
quantized channel gains are typical. 

We can then assume that every matrix of quantized channel gains
$\hat{\bm{\msf{H}}}$ appears exactly\footnote{Since $T/L$ will grow to
infinity, we can assume here that \eqref{eq:frac} is integer and avoid
floor operators.}
\begin{equation}
    \label{eq:frac}
    (1-\eta) p_{\hat{\bm{H}}}( \hat{\bm{\msf{H}}} )T/L
\end{equation}
many times in each of the $L$ blocks, ignoring all the remaining time
slots. This results in a loss of at most a factor $(1-\eta)$ in rate.
Furthermore, we may assume without loss of generality that the first
$(1-\eta)T/L$ quantized channel gains in each subblock satisfy this
condition.

We now describe the matching procedure alluded to earlier.  Consider the
channel gains at time $t_1=1$ in the first of the $L$ subblocks and the
corresponding matrix of quantized channel gains $\hat{\bm{H}}[t_1]$. Let
$t_\ell$ be the first time in subblock $\ell\in\{2,\cdots,L\}$ such that
\begin{subequations}
    \label{eq:match}
    \begin{align}
        \label{eq:match1}
        \hat{h}_{1,1}[t_\ell] & = \hat{h}_{1,1}[t_1], \\
        \hat{h}_{1,2}[t_\ell] & = \hat{h}_{1,2}[t_1], \\
        \hat{h}_{2,2}[t_\ell] & = \hat{h}_{2,2}[t_1], \\
        \label{eq:match4}
        \hat{h}_{2,1}[t_\ell] & = \omega_L^{\ell-1} \hat{h}_{2,1}[t_1],
    \end{align}
\end{subequations}
where
\begin{equation*}
    \omega_L \defeq \exp\big(\sqrt{-1}\tfrac{2\pi}{L}\big)
\end{equation*}
is the $L$th root of unity. By construction of the quantization scheme,
if $\hat{h}\in\hat{\mc{H}}$ then
$\omega_L^{\ell-1}\hat{h}\in\hat{\mc{H}}$, and hence such a collection
of time slots $t_2, \dots, t_L$ can exist. Since $t_1 < t_2 < \dots <
t_L$, this matching procedure can be performed in a causal manner and
using only instantaneous CSI.  Moreover, by the full CSI assumption,
this matching can be computed at each transmitter and receiver. Note
that, as discussed in the motivating example in
Section~\ref{sec:motivate}, the choice of $\hat{h}_{2,1}$ is used to
shift the symbol pairings at the second receiver. This in turn makes it
possible to create orthogonal integer-valued subchannels at both
receivers via careful power allocation.   

Having performed the matching for $t_1=1$, we proceed with $t_1 = 2$.
We again match channel gains in the same fashion, ensuring that each
time slot $t_\ell$ in subblock $\ell\in\{2,\ldots, L\}$ is chosen at
most once. In other words, this matching procedure constructs many 
nonintersecting $L$-element subsets $\{t_1, \ldots, t_L\}$ of
$\{1,\ldots, T\}$. We now argue that this procedure can be continued
successfully up to $t_1 = (1-\eta)T/L$, i.e., $(1-\eta)T/L$ of these
subsets can be found.

Consider a time slot $t_1$ in the first subblock and the corresponding
channel gains $\hat{\bm{H}}[t_1]$. This channel gain induces matched
channel gains 
\begin{equation*}
    \hat{\bm{H}}[t_2], 
    \hat{\bm{H}}[t_3], 
    \ldots,
    \hat{\bm{H}}[t_L], 
\end{equation*}
within subblocks $2, \ldots, L$. Hence, the distribution of the channel
gains $\hat{\bm{H}}[t_1]$ at some fixed $t_1$ induces a distribution of
the channel gains $\hat{\bm{H}}[t_\ell]$ for $\ell\in\{2,\ldots, L\}$.
It is not clear \emph{a priori} that $\hat{\bm{H}}[t_\ell]$ and
$\hat{\bm{H}}[t]$ for any fixed $t$ have the same distribution.

The key observation for the analysis of the matching procedure is the
following. By \eqref{eq:frac}, the matching procedure is successful for
all $t_1\in\{1,\ldots, (1-\eta)T/L\}$ if the distribution of
$\hat{\bm{H}}[t_\ell]$ for $\ell\in\{2\ldots, L\}$ is the same as the
distribution of $\hat{\bm{H}}[(\ell-1) T/L+1]$ (or any other channel
matrix at \emph{fixed} time $t$ in subblock $\ell$). By stationarity,
the distribution of $\hat{\bm{H}}[(\ell-1) T/L+1]$ is the same as the
distribution of $\hat{\bm{H}}[1]$. Hence, it suffices to argue that
$\hat{\bm{H}}[t_\ell]$ has the same distribution as $\hat{\bm{H}}[1]$,
i.e., that $\hat{\bm{H}}[t_\ell]$ has distribution $p_{\hat{\bm{H}}}$.
We now show that this is the case.

By assumption, the distribution of each channel gain $h_{m,k}[t]$ is
circularly symmetric. By Lemma~\ref{thm:symmetry}, the quantization
scheme preserves this circular symmetry, in the sense that all possible
quantized channel gains with the same magnitude have the same
probability. Since the components of $\hat{\bm{H}}[t]$ are independent
by Lemma~\ref{thm:symmetry}, this circular symmetry also holds for their
joint distribution, i.e., if $\hat{\bm{\msf{H}}}$ and
$\smash{\hat{\bm{\msf{H}}}}'$ satisfy $\abs{\hat{\msf{h}}_{m,k}} =
\abs{\hat{\msf{h}}_{m,k}'}$ for all $m,k$, then 
\begin{equation*}
    p_{\hat{\bm{H}}}(\smash{\hat{\bm{\msf{H}}}}')
    = p_{\hat{\bm{H}}}(\hat{\bm{\msf{H}}}).
\end{equation*}

Observe now that, for each $m,k$, the channel gains 
\begin{equation*}
    \hat{h}_{m,k}[t_1], 
    \hat{h}_{m,k}[t_2], 
    \ldots,
    \hat{h}_{m,k}[t_L], 
\end{equation*}
all have the same magnitude by the matching condition \eqref{eq:match}.
Moreover, since the distribution of $\hat{\bm{H}}[t_1]$ is circularly
symmetric, and since \eqref{eq:match} results in a \emph{fixed} phase
shift, the induced distribution of the matched channel gains
$\hat{\bm{H}}[t_\ell]$ is circularly symmetric as well. Together, these
two facts show that the distribution of the quantized channel gains
induced by the matching within the subblocks $\ell\in\{2,\ldots, L\}$ is
identical to the distribution of the quantized channel gains within the
first subblock.  This implies that the time slots $t_1=1$ up to
$t_1=(1-\eta)T/L$ can be matched by the described procedure.

Out of the $(1-\eta)T/L$ time slots that are matched in this fashion, at
most $\delta T/L$ contain a quantized channel gain equal to
infinity by \eqref{eq:infty} for some $\delta = \delta(\nu)$ (where
$\nu$ is the parameter governing the number of quantization points).
These time slots are not used. Again by the full CSI assumption, this
event can be observed at each transmitter and receiver. Accounting for
the time slots that are not matched, a total of at least
$(1-\eta-\delta)T/L$ time slots in each subblock are used for
communication.

To summarize, the channel gains in each of the $L$ subblocks are matched
up to satisfy \eqref{eq:match}. With probability at least 
\begin{equation}
    \label{eq:error}
    1-\varepsilon_1(T),
\end{equation}
at least a fraction 
\begin{equation*}
    (1-\eta(T)-\delta(\nu))
\end{equation*}
of the time slots in each subblock can be matched in this fashion such
that all the corresponding channel gains have finite magnitudes. Here
the parameters can be chosen to satisfy
\begin{align}
    \label{eq:epsilon1}
    \lim_{T\to\infty} \varepsilon_1(T) & = 0, \\
    \label{eq:eta}
    \lim_{T\to\infty} \eta(T) & = 0, 
\end{align}
both for fixed values of $L$ and $\nu$, and 
\begin{equation}
    \label{eq:delta}
    \lim_{\nu\to\infty} \delta(\nu)  = 0.
\end{equation}

\subsection{Precoding and Matched Filtering}
\label{sec:proofs_K2_precoding}

Consider time slots $t_1, \ldots, t_L$ in subblocks $1, \ldots,
L$ that are matched as described in the last section. We now describe a
linear precoding transmitter design and matched filtering receiver
design that transform the complex channel over these $L$ time slots
into parallel integer channels.

Construct the diagonal matrix 
\begin{equation*}
    \bm{D}_{m,k} 
    \defeq \diag\big((h_{m,k}[t_\ell])_{\ell=1}^L\big),
\end{equation*}
from the $L$ matched channel gains between transmitter $k$ and receiver
$m$ and define $\hat{\bm{D}}_{m,k}$ in the same manner, but with respect to
$\hat{h}_{m,k}[t_\ell]$. 
Observe from \eqref{eq:match} that
\begin{equation*}
    \hat{\bm{D}}_{m,k} 
    = 
    \begin{cases}
        \hat{h}_{m,k}\bm{I}, & \text{if $(m,k)\neq(2,1)$} \\
        \hat{h}_{m,k}\bm{F}, & \text{if $(m,k)=(2,1)$}
    \end{cases}
\end{equation*}
by the matching procedure, where 
\begin{equation*}
    \hat{h}_{m,k} \defeq \hat{h}_{m,k}[t_1]
\end{equation*}
and 
\begin{equation*}
    \bm{F} \defeq \diag\big((\omega_L^{\ell-1})_{\ell=1}^{L}\big).
\end{equation*}

Denote by 
\begin{equation*}
    \bm{x}_k\defeq
    \begin{pmatrix}
        x_k[t_1] & x_k[t_2] & \dots & x_k[t_L]
    \end{pmatrix}^\T
\end{equation*}
the vector of channel inputs at time slots $t_1, \ldots, t_L$ at
transmitter $k\in\{1,2\}$. Similarly, denote by 
\begin{equation*}
    \bm{y}_m\defeq
    \begin{pmatrix}
        y_m[t_1] & y_m[t_2] & \dots & y_m[t_L]
    \end{pmatrix}^\T
\end{equation*}
and
\begin{equation*}
    \bm{z}_m\defeq
    \begin{pmatrix}
        z_m[t_1] & z_m[t_2] & \dots & z_m[t_L]
    \end{pmatrix}^\T
\end{equation*}
the vector of channel outputs and noises at time slots $t_1, \ldots,
t_L$ at receiver $m\in\{1,2\}$. The relationship between $\bm{x}_k$ and
$\bm{y}_m$ is given by
\begin{equation}
    \label{eq:block}
    \bm{y}_m = \bm{D}_{m,1}\bm{x}_1+\bm{D}_{m,2}\bm{x}_2+\bm{z}_m
\end{equation}
for $m\in\{1,2\}$. 

Each transmitter uses a linear precoder over the block
channel~\eqref{eq:block}. Transmitter one has access to $L$ symbols
$s_{1,1}, \ldots, s_{1,L}$ and transmitter two has access to $L-1$
symbols $s_{2,1},\ldots, s_{2,L-1}$. We assume that all these $2L-1$
symbols have zero mean and are mutually independent. We will provide a
detailed description as to how these symbols constitute codewords across
matchings of time slots in Section \ref{sec:proofs_K2_computation}. Each
message symbol is multiplied by a transmit vector in $\C^L$. Transmitter
one uses a total of $L$ transmit vectors $\bm{v}_{1,1},\ldots,
\bm{v}_{1,L} \in \C^L$ and transmitter two uses $L-1$ transmit vectors
$\bm{v}_{2,1},\ldots, \bm{v}_{2,L-1} \in \C^L$. The modulated transmit
vectors are summed up by the transmitter, and, at time $t_\ell$, the
$\ell$th component of this sum of vectors is sent over the channel. The
resulting channel input vector $\bm{x}_k$ at transmitter $k\in\{1,2\}$
is given by
\begin{subequations}
    \label{eq:xdef}
    \begin{align}
        \bm{x}_1 & = \sum_{\ell=1}^L s_{1,\ell}\bm{v}_{1,\ell} \\
        \shortintertext{and} \nonumber\\
        \bm{x}_2 & = \sum_{\ell=1}^{L-1} s_{2,\ell}\bm{v}_{2,\ell}.
    \end{align}
\end{subequations}
Substituting \eqref{eq:xdef} into \eqref{eq:block} yields
\begin{subequations}
    \label{eq:subchannels}
    \begin{align}
        \bm{y}_1 
        & = (s_{1,1} \bm{D}_{1,1} \bm{v}_{1,1} + s_{2,1} \bm{D}_{1,2} \bm{v}_{2,1}) 
        + (s_{1,2} \bm{D}_{1,1} \bm{v}_{1,2} + s_{2,2} \bm{D}_{1,2} \bm{v}_{2,2}) + \cdots \nonumber\\
        & \quad {} +(s_{1,L-1} \bm{D}_{1,1} \bm{v}_{1,L-1} + s_{2,L-1} \bm{D}_{12} \bm{v}_{2,L-1}) 
        + s_{1,L} \bm{D}_{1,1} \bm{v}_{1,L} + \bm{z}_1 \\
        \shortintertext{and}\nonumber\\
        \bm{y}_2 & = 
        (s_{1,2} \bm{D}_{2,1} \bm{v}_{1,2} + s_{2,1} \bm{D}_{2,2} \bm{v}_{2,1}) 
        + (s_{1,3} \bm{D}_{2,1} \bm{v}_{1,3} + s_{2,2} \bm{D}_{2,2} \bm{v}_{2,2}) + \cdots \nonumber\\
        & \quad {} +(s_{1,L} \bm{D}_{2,1} \bm{v}_{1,L} + s_{2,L-1} \bm{D}_{2,2} \bm{v}_{2,L-1}) 
        + s_{1,1} \bm{D}_{2,1} \bm{v}_{1,1} + \bm{z}_2.
    \end{align} 
\end{subequations}

Our goal is to create $L$ orthogonal subchannels, indicated by the
parentheses in \eqref{eq:subchannels}, with integer-valued coefficients
at each receiver. We now demonstrate how this can be achieved through an
appropriate choice of transmit vectors.  Consider first the special case
where the channel coefficients all have unit magnitudes, i.e.,
\mbox{$\abs{h_{m,k}} = 1$} for all $m,k$.  Assume the transmit vectors
$\bm{v}_{k,\ell}$ satisfy the following four \emph{computation-alignment
conditions}:
\begin{enumerate}
    \item {$\bm{D}_{1,1} \bm{v}_{1,\ell} = \bm{D}_{1,2}\bm{v}_{2,\ell}$,
        for $\ell\in\{1,\ldots,L-1\}$;}
    \item {$\bm{D}_{2,1}\bm{v}_{1,\ell} =
        \bm{D}_{2,2}\bm{v}_{2,\ell-1}$, for $\ell\in\{2,3,\ldots, L\}$;}
    \item {$\{\bm{D}_{1,1}\bm{v}_{1,1},\ldots ,
        \bm{D}_{1,1}\bm{v}_{1,L}\}$ are orthogonal to each other;} 
    \item {$\{\bm{D}_{2,1}\bm{v}_{1,1}, \ldots ,
        \bm{D}_{2,1}\bm{v}_{1,L}\}$ are orthogonal to each other.} 
\end{enumerate} 
Then, by the first and second alignment conditions,
\eqref{eq:subchannels} can be rewritten as
\begin{align*}
    \bm{y}_1 
    & = (s_{1,1}+s_{2,1})\bm{D}_{1,1} \bm{v}_{1,1}
    + (s_{1,2} + s_{2,2})\bm{D}_{1,1} \bm{v}_{1,2} + \cdots \\
    & \quad {} +(s_{1,L-1} + s_{2,L-1}) \bm{D}_{1,1} \bm{v}_{1,L-1} 
    + s_{1,L} \bm{D}_{1,1} \bm{v}_{1,L} + \bm{z}_1 \\
    \bm{y}_2 & = 
    (s_{1,2} + s_{2,1})\bm{D}_{2,1} \bm{v}_{1,2} 
    + (s_{1,3} + s_{2,2}) \bm{D}_{2,1} \bm{v}_{1,3} + \cdots \\
    & \quad {} +(s_{1,L}+ s_{2,L-1})\bm{D}_{2,1} \bm{v}_{1,L} 
    + s_{1,1} \bm{D}_{2,1} \bm{v}_{1,1} + \bm{z}_2.
\end{align*} 
Note that each subchannel consists of the sum of two symbols
$s_{k,\ell}$ multiplied by some vector $\bm{D}_{1,1} \bm{v}_{1,\ell}$ or
$\bm{D}_{2,1} \bm{v}_{1,\ell}$. By the third and fourth alignment
conditions, these vectors are orthogonal and can hence be recovered
without any interference using matched filters at the receiver. Thus, we
have transformed the channel with complex channel coefficients into several
orthogonal subchannels with integer channel coefficients over which
lattice codes can be efficiently used.

For arbitrary channel matrices $\bm{D}_{m,k}$, satisfying the
computation-alignment conditions is not possible. However, we now argue
that due to the special form of $\bm{D}_{m,k}$ resulting from the
matching procedure described in Section~\ref{sec:proofs_K2_matching},
this is possible here. Assume for the moment that the channel gains
$\bm{D}_{m,k}$ are equal to their quantized version
$\hat{\bm{D}}_{m,k}$. Then it can be verified that the following choice
of the transmit vectors satisfies the computation-alignment conditions:
\begin{align*}
    \bm{v}_{1,1} 
    & = (1 \  1 \  \dots \  1)^\T \\
    \bm{v}_{1,\ell}
    & = \hat{\bm{D}}_{2,1}^{-1} \hat{\bm{D}}_{2,2} 
    \hat{\bm{D}}_{1,2}^{-1} \hat{\bm{D}}_{1,1} \bm{v}_{1,\ell-1} 
    = \frac{\hat{h}_{2,2}\hat{h}_{1,1}}{\hat{h}_{2,1}\hat{h}_{1,2}}\bm{F}^{-1}\bm{v}_{1,\ell-1}, 
    \quad \ell\in\{2,3,\ldots,L\} \\
    \bm{v}_{2,\ell} 
    & = \hat{\bm{D}}_{1,2}^{-1} \hat{\bm{D}}_{1,1} \bm{v}_{1,\ell} 
    = \frac{\hat{h}_{1,1}}{\hat{h}_{1,2}}\bm{v}_{1,\ell}, \quad \ell\in\{1,\ldots, L\}.
\end{align*} 

Turning to the case with general channel magnitudes $\abs{h_{m,k}}$, we
observe that this recursive construction leads to transmit vectors with
exponentially different norms as $L$ increases, i.e.,
\begin{equation*}
    \norm{\bm{v}_{1,L}}  
    = \biggl(\frac{\abs{\hat{h}_{2,2}}\abs{\hat{h}_{1,1}}}
    {\abs{\hat{h}_{2,1}}\abs{\hat{h}_{1,2}}}\biggr)^{L-1}
    \norm{\bm{v}_{1,1}}. 
\end{equation*}
This causes extremely unequal power allocation across the transmit
vectors for large $L$, resulting in a significant rate loss and
precluding a constant-gap capacity approximation. To circumvent this
issue, we will relax the computation-alignment condition, which in turn
will allow us to equalize the vector lengths using a scaling factor. 

Observe that the first and second computation-alignment conditions
guarantee that each of the orthogonal subchannels carries the sum of two
signals. This is sufficient for the efficient use of lattice codes, but
not necessary. Indeed a weaker sufficient condition is that each of the
orthogonal subchannels carries an integer linear combination of the
signals. We can thus relax the second computation-alignment condition to
\begin{enumerate}
    \item[2')] {$\bm{D}_{2,1}\bm{v}_{1,\ell} =
        b_j\bm{D}_{2,2}\bm{v}_{2,\ell-1}$, for $\ell\in\{2,3,\ldots, L\}$}
\end{enumerate}
where the scalar $b_j$ is an integer or its inverse.

These relaxed conditions are satisfied by
\begin{subequations}
    \label{eq:vectors}
    \begin{align}
        \bm{v}_{1,1} 
        & = (1 \ 1 \ \dots \ 1)^\T \\
        \bm{v}_{1,\ell}
        & = b_{\ell} \hat{\bm{D}}_{2,1}^{-1} \hat{\bm{D}}_{2,2} 
        \hat{\bm{D}}_{1,2}^{-1} \hat{\bm{D}}_{1,1} \bm{v}_{1,\ell-1} 
        = b_{\ell} \frac{\hat{h}_{2,2}\hat{h}_{1,1}}{\hat{h}_{2,1}\hat{h}_{1,2}}
        \bm{F}^{-1}\bm{v}_{1,\ell-1}, \quad\ell\in\{2,3,\ldots,L\} \\
        \bm{v}_{2,\ell}
        & = \hat{\bm{D}}_{1,2}^{-1} \hat{\bm{D}}_{1,1} \bm{v}_{1,\ell} 
        = \frac{\hat{h}_{1,1}}{\hat{h}_{1,2}}\bm{v}_{1,\ell}, \quad \ell\in\{1,\ldots, L\}.
    \end{align} 
\end{subequations}
where the scalar $b_\ell$ is of the form $n$ or $1/n$ for the smallest
natural number $n\in\N$ such that 
\begin{equation}
    \label{eq:cbound2}
    \norm{\bm{v}_{1,\ell}}/\sqrt{L}\in[1,2).
\end{equation}
For convenience of notation, we set $b_1\defeq 1$. Note that scalar
$b_\ell$ equalizes all transmit vectors to have approximately the same
norm, as desired.

We now analyze the performance of this choice of transmit vectors in
detail. Define
\begin{equation}
    \label{eq:cdef}
    c 
    = c(\hat{\bm{H}})
    \defeq \prod_{m,k}
    \max\big\{\abs{\hat{h}_{m,k}}, \abs{\hat{h}_{m,k}}^{-1}\big\}.
\end{equation}
It follows from \eqref{eq:vectors} and \eqref{eq:cbound2} that
\begin{equation}
    \label{eq:cbound}
    1/c 
    \leq \frac{\abs{\hat{h}_{1,1}}}{\abs{\hat{h}_{1,2}}}
    \leq \norm{\bm{v}_{2,\ell}}/\sqrt{L}
    \leq 2 \frac{\abs{\hat{h}_{1,1}}}{\abs{\hat{h}_{1,2}}}
    \leq 2c
\end{equation}
and that
\begin{equation}
    \label{eq:bbound}
    \max\{b_\ell, b_\ell^{-1}\} \leq 2c.
\end{equation}

We allocate the same amount of power 
\begin{equation}
    \label{eq:psub}
    \E\big(\abs{s_{k,\ell}}^2\big)
    = \frac{P}{4Lc^2}
    \defeq \tilde{P} 
\end{equation} to each symbol $s_{k,\ell}$. Since
$\norm{\bm{v}_{k,\ell}}^2\leq 4Lc^2$ by \eqref{eq:cbound2} and
\eqref{eq:cbound}, we have using the construction of $\bm{x}_k$ in
\eqref{eq:xdef},
\begin{equation*}
    \frac{1}{L}\E\big(\norm{\bm{x}_k}^2\big) 
    \leq P,
\end{equation*} satisfying the overall average power constraint of $P$ over the $L$ time
slots $t_1, \ldots, t_L$. 

The operation of the receivers is implemented by multiplying
the vector of channel outputs $\bm{y}_m$ by the matched filter
\begin{equation}
    \label{eq:vtildedef}
    \tilde{\bm{v}}_{m,j} 
    \defeq 
    \bm{v}_{m,j}/\norm{\bm{v}_{m,j}}
\end{equation}
for $m=1, j\in\{1,\ldots, L\}$ and for $m=2, j\in\{1,\ldots, L-1\}$,
to form
\begin{equation*}
    \tilde{\bm{v}}_{m,j}^\dagger \bm{y}_m 
    = \sum_{\ell=1}^L s_{1,\ell}
    \tilde{\bm{v}}_{m,j}^\dagger\bm{D}_{m,1}\bm{v}_{1,\ell} 
    +\sum_{\ell=1}^{L-1} s_{2,\ell}
    \tilde{\bm{v}}_{m,j}^\dagger\bm{D}_{m,2}\bm{v}_{2,\ell}
    + \tilde{\bm{v}}_{m,j}^\dagger\bm{z}_m.
\end{equation*}

In general, the channel gains are not equal to their quantized versions,
i.e., $\bm{D}_{m,k} \neq \hat{\bm{D}}_{m,k}$. However, since we only
communicate during time slots satisfying $\abs{h_{m,k}[t]}\leq \nu$, the
quantization error is upper bounded by \eqref{eq:quant} as
\begin{equation*}
    \abs{h_{m,k}[t_\ell]-\hat{h}_{m,k}[t_\ell]} \leq (\pi+1)/\nu,
\end{equation*}
so the matrices $\bm{D}_{m,k}$ and $\hat{\bm{D}}_{m,k}$ are quite close
for quantization parameter $\nu$ large enough. We will use the same
transmitter and receiver structures as for the perfectly matched case,
i.e., \eqref{eq:vectors} and \eqref{eq:vtildedef}. The
computation-alignment conditions are then only approximately satisfied.
To determine performance, we will bound the additional interference that
is caused by imperfect alignment (received vectors do not line up) and
imperfect zero forcing of interference (received vectors are not
orthogonal). 

Define
\begin{equation*}
    \bm{\Upsilon}_{m,k} \defeq \bm{D}_{m,k}-\hat{\bm{D}}_{m,k}
\end{equation*} 
as the (diagonal) matrix of channel quantization errors. We can then
rewrite the output of the matched filter at receiver one as
\begin{subequations}
    \label{eq:3parts}
    \begin{align}
        \label{eq:3parts_1}
        \tilde{\bm{v}}_{1,j}^\dagger \bm{y}_1
        & =
        \bigl(s_{1,j} \tilde{\bm{v}}_{1,j}^\dagger \hat{\bm{D}}_{1,1}\bm{v}_{1,j}
        + s_{2,j} \tilde{\bm{v}}_{1,j}^\dagger \hat{\bm{D}}_{1,2}\bm{v}_{2,j}\bigr)
        \nonumber\\
        & \quad {}+
        \Bigl(s_{1,j} \tilde{\bm{v}}_{1,j}^\dagger \bm{\Upsilon}_{1,1}\bm{v}_{1,j}
        + s_{2,j} \tilde{\bm{v}}_{1,j}^\dagger \bm{\Upsilon}_{1,2}\bm{v}_{2,j} 
        +
        {\textstyle \sum_{\ell\neq j}} s_{1,\ell}
        \tilde{\bm{v}}_{1,j}^\dagger
        \bm{D}_{1,1}\bm{v}_{1,\ell}
        +{\textstyle\sum_{\ell\neq j}} s_{2,\ell}
        \tilde{\bm{v}}_{1,j}^\dagger
        \bm{D}_{1,2}\bm{v}_{2,\ell}\Bigr) \nonumber\\
        & \quad {}+
        \tilde{\bm{v}}_{1,j}^\dagger\bm{z}_1
    \end{align}
    for $j\in\{1,\ldots, L-1\}$ and as
    \begin{align}
        \label{eq:3parts_2}
        \tilde{\bm{v}}_{1,L}^\dagger \bm{y}_1
        & =
        s_{1,L} \tilde{\bm{v}}_{1,L}^\dagger \hat{\bm{D}}_{1,1}\bm{v}_{1,L}
        \nonumber\\
        & \quad {}+
        \Bigl(s_{1,L} \tilde{\bm{v}}_{1,L}^\dagger \bm{\Upsilon}_{1,1}\bm{v}_{1,L}
        + {\textstyle\sum_{\ell=1}^{L-1}} s_{1,\ell}
        \tilde{\bm{v}}_{1,L}^\dagger
        \bm{D}_{1,1}\bm{v}_{1,\ell} 
        +{\textstyle\sum_{\ell=1}^{L-1}} s_{2,\ell}
        \tilde{\bm{v}}_{1,L}^\dagger
        \bm{D}_{1,2}\bm{v}_{2,\ell}\Bigr) \nonumber\\
        & \quad {}+
        \tilde{\bm{v}}_{1,L}^\dagger\bm{z}_1
    \end{align}
    for $j=L$. Similarly, we can rewrite the output of the matched filter at
    receiver two as
    \begin{align}
        \label{eq:3parts_3}
        \tilde{\bm{v}}_{2,j}^\dagger \bm{y}_2 
        & =
        \bigl(s_{1,j+1} \tilde{\bm{v}}_{2,j}^\dagger \hat{\bm{D}}_{2,1}\bm{v}_{1,j+1}
        + s_{2,j} \tilde{\bm{v}}_{2,j}^\dagger \hat{\bm{D}}_{2,2}\bm{v}_{2,j}\bigr)
        \nonumber\\
        & \quad {}+
        \Bigl(s_{1,j+1} \tilde{\bm{v}}_{2,j}^\dagger \bm{\Upsilon}_{2,1}\bm{v}_{1,j+1}
        + s_{2,j} \tilde{\bm{v}}_{2,j}^\dagger \bm{\Upsilon}_{2,2}\bm{v}_{2,j} 
        + {\textstyle\sum_{\ell\neq j+1}} s_{1,\ell}
        \tilde{\bm{v}}_{2,j}^\dagger
        \bm{D}_{2,1}\bm{v}_{1,\ell}
        +{\textstyle\sum_{\ell\neq j}} s_{2,\ell}
        \tilde{\bm{v}}_{2,j}^\dagger
        \bm{D}_{2,2}\bm{v}_{2,\ell}\Bigr) \nonumber\\
        & \quad {}+
        \tilde{\bm{v}}_{2,j}^\dagger\bm{z}_2
    \end{align}
\end{subequations}
for $j\in\{1,\ldots, L-1\}$. From
\eqref{eq:3parts}, we see that the matched filter
output consists of three parts: desired signal, mismatch terms due to
imperfect alignment and imperfect zero forcing of interference, and
receiver noise.

We start with the analysis of the desired signals in \eqref{eq:3parts}.
The desired signal at receiver one is
\begin{subequations}
    \label{eq:desired}
    \begin{equation}
        \label{eq:desired1}
        s_{1,j} \tilde{\bm{v}}_{1,j}^\dagger\hat{\bm{D}}_{1,1}\bm{v}_{1,j}
        + s_{2,j}\tilde{\bm{v}}_{1,j}^\dagger\hat{\bm{D}}_{1,2}\bm{v}_{2,j} 
        = \hat{h}_{1,1} \norm{\bm{v}_{1,j}} (s_{1,j} + s_{2,j})  
    \end{equation}
    for $j\in\{1,\ldots, L-1\}$ and
    \begin{equation}
        \label{eq:desired2}
        s_{1,L} \tilde{\bm{v}}_{1,L}^\dagger\hat{\bm{D}}_{1,1}\bm{v}_{1,L}
        = \hat{h}_{1,1} \norm{\bm{v}_{1,L}} s_{1,L} 
    \end{equation}
    for $j=L$, where we have used \eqref{eq:vectors} and
    \eqref{eq:vtildedef}.  Similarly, the desired signal at receiver two
    is
    \begin{equation}
        \label{eq:desired3}
        s_{1,j+1} \tilde{\bm{v}}_{2,j}^\dagger\hat{\bm{D}}_{2,1}\bm{v}_{1,j+1}
        + s_{2,j}\tilde{\bm{v}}_{2,j}^\dagger\hat{\bm{D}}_{2,2}\bm{v}_{2,j} 
        = \hat{h}_{2,2} \norm{\bm{v}_{2,j}} (b_{j+1}s_{1,j+1} + s_{2,j}) 
    \end{equation}
\end{subequations}
for $j\in\{1,\ldots, L-1\}$. The received signal power (for each symbol)
satisfies
\begin{subequations}
    \label{eq:power}
    \begin{align}
        \label{eq:power1}
        \abs{\hat{h}_{1,1}}^2 \norm{\bm{v}_{1,j}}^2 \E\big(\abs{s_{k,j}}^2 \big) 
        \stackrel{(a)}{\geq}\abs{\hat{h}_{1,1}}^2 L \tilde{P}
        \stackrel{(b)}{\geq} \frac{L\tilde{P}}{c^2}
    \end{align} 
    at receiver one, where we have used~\eqref{eq:cbound2}
    and~\eqref{eq:psub} in $(a)$ and~\eqref{eq:cdef} in $(b)$.
    Similarly, using \eqref{eq:cbound} instead of~\eqref{eq:cbound2},
    \begin{align}
        \label{eq:power2}
        \abs{\hat{h}_{2,2}}^2 \norm{\bm{v}_{2,j}}^2 \E\big(\abs{s_{k,j}}^2 \big) 
        \geq \frac{\abs{\hat{h}_{2,2}}^2 \abs{\hat{h}_{1,1}}^2}{\abs{\hat{h}_{1,2}}^2} 
        L \tilde{P}
        \geq \frac{L\tilde{P}}{c^2}
    \end{align} 
\end{subequations}
at receiver two (not accounting for the normalization factor $b_{j+1}$).

Before we continue with the analysis of the mismatch terms in
\eqref{eq:3parts}, we argue
that $\abs{ \tilde{\bm{v}}_{m,j}^\dagger \bm{\Upsilon}_{m,k}
\bm{v}_{k,\ell} }^2$ is small.  By the Cauchy-Schwarz inequality,
\begin{equation}
    \label{eq:residual1}
    \abs{ \tilde{\bm{v}}_{m,j}^\dagger \bm{\Upsilon}_{m,k} \bm{v}_{k,\ell} }^2 
    \leq  \norm{\tilde{\bm{v}}_{m,j}}^2 \norm{\bm{\Upsilon}_{m,k}}^2 
    \norm{\bm{v}_{k,\ell}}^2,
\end{equation} 
where $\norm{\bm{\Upsilon}_{m,k}}^2$ denotes the sum of squared diagonal
entries of $\bm{\Upsilon}_{m,k}$.  By construction,
$\norm{\tilde{\bm{v}}_{k,j}}^2 = 1$.  From \eqref{eq:quant},
$\norm{\bm{\Upsilon}_{m,k}}^2$ satisfies 
\begin{equation*}
    \norm{\bm{\Upsilon}_{m,k}}^2 \leq L (\pi +1)^2/\nu^2.
\end{equation*}
By \eqref{eq:cbound2} and \eqref{eq:cbound},
\begin{equation*}
    \norm{\bm{v}_{k,j}}^2 \leq 4Lc^2
\end{equation*}
for $k\in\{1,2\}$, where we have used that $c \geq 1$ by
\eqref{eq:cdef}. Combining this with \eqref{eq:residual1} yields the
desired upper bound
\begin{equation}
    \label{eq:residual}
    \abs{ \tilde{\bm{v}}_{m,j}^\dagger \bm{\Upsilon}_{m,k} \bm{v}_{k,\ell} }^2 
    \leq \frac{4 L^2 (\pi + 1)^2c^2}{\nu^2}
    \defeq \gamma^2.
\end{equation} 

The mismatch term in \eqref{eq:3parts} due to
imperfect alignment is
\begin{subequations}
    \label{eq:mismatch_alignment}
    \begin{equation}
        \label{eq:mismatch_alignment1}
        s_{1,j} \tilde{\bm{v}}_{1,j}^\dagger\bm{\Upsilon}_{1,1}\bm{v}_{1,j}
        + s_{2,j} \tilde{\bm{v}}_{1,j}^\dagger\bm{\Upsilon}_{1,2}\bm{v}_{2,j} 
        \defeq e_{1,1,j} s_{1,j} + e_{1,2,j} s_{2,j} 
    \end{equation} 
    at receiver one, and 
    \begin{equation}
        \label{eq:mismatch_alignment2}
        s_{1,j+1} \tilde{\bm{v}}_{2,j}^\dagger\bm{\Upsilon}_{2,1}\bm{v}_{1,j+1}
        +s_{2,j} \tilde{\bm{v}}_{2,j}^\dagger\bm{\Upsilon}_{2,2}\bm{v}_{2,j} 
        \defeq e_{2,1,j} s_{1,j+1} + e_{2,2,j} s_{2,j}
    \end{equation}
\end{subequations}
at receiver two. Each term $e_{m,k,j}$ can be interpreted as the residual
channel fluctuation after the quantized matching, and satisfies
\begin{equation}
    \label{eq:gamma}
    \abs{e_{m,k,j}}^2 \leq \gamma^2
\end{equation}
by \eqref{eq:residual}.

The mismatch term in \eqref{eq:3parts} due to
imperfect zero forcing is
\begin{subequations}
    \label{eq:mismatch_zero}
    \begin{align}
        \label{eq:mismatch_zero1}
        \theta_{1,j} 
        & \defeq \sum_{\ell\neq j} s_{1,\ell} 
        \tilde{\bm{v}}_{1,j}^\dagger\bm{D}_{1,1}\bm{v}_{1,\ell}
        +\sum_{\ell\neq j} s_{2,\ell} 
        \tilde{\bm{v}}_{1,j}^\dagger\bm{D}_{1,2}\bm{v}_{2,\ell} \nonumber\\
        & = \sum_{\ell\neq j} s_{1,\ell} 
        \tilde{\bm{v}}_{1,j}^\dagger\bm{\Upsilon}_{1,1}\bm{v}_{1,\ell}
        +\sum_{\ell\neq j} s_{2,\ell} 
        \tilde{\bm{v}}_{1,j}^\dagger\bm{\Upsilon}_{1,2}\bm{v}_{2,\ell}
    \end{align}
    at receiver one, where we have used the orthogonality of the received
    vectors under channel gains $\hat{\bm{D}}_{m,k}$. Similarly, 
    \begin{align}
        \label{eq:mismatch_zero2}
        \theta_{2,j}
        & \defeq \sum_{\ell\neq j+1} s_{1,\ell} 
        \tilde{\bm{v}}_{2,j}^\dagger\bm{D}_{2,1}\bm{v}_{2,\ell}
        +\sum_{\ell\neq j} s_{2,\ell} 
        \tilde{\bm{v}}_{2,j}^\dagger\bm{D}_{2,2}\bm{v}_{2,\ell} \nonumber\\
        & = \sum_{\ell\neq j+1} s_{1,\ell} 
        \tilde{\bm{v}}_{2,j}^\dagger\bm{\Upsilon}_{2,1}\bm{v}_{2,\ell}
        +\sum_{\ell\neq j} s_{2,\ell} 
        \tilde{\bm{v}}_{2,j}^\dagger\bm{\Upsilon}_{2,2}\bm{v}_{2,\ell}
    \end{align}
\end{subequations}
at receiver two. Using \eqref{eq:psub} and \eqref{eq:residual} together
with the independence of the signals $s_{k,\ell}$, the
total zero-forcing leakage power 
\begin{align}
    \label{eq:sigmadef}
    \sigma^2 
    \defeq \max_{m,j} \E\big(\abs{\theta_{m,j}}^2\big) 
\end{align} 
is upper bounded by 
\begin{equation}
    \label{eq:sigmabound}
    \sigma^2 \leq 2 (L-1) \gamma^2 \tilde{P}  
\end{equation}
at each receiver.

Finally, the additive noise term 
\begin{equation}
    \label{eq:noise}
    \tilde{z}_{m,j} 
    \defeq \tilde{\bm{v}}_{m,j}^\dagger\bm{z}_m
\end{equation}
in \eqref{eq:3parts}
is circularly-symmetric complex Gaussian with mean zero and variance
one, since $\norm{\tilde{\bm{v}}_{m,j}}^2=1$. 

Substituting \eqref{eq:desired}, \eqref{eq:mismatch_alignment},
\eqref{eq:mismatch_zero}, and \eqref{eq:noise} into \eqref{eq:3parts},
yields that the output of the $j$th matched filter at receiver one is
\begin{equation}
    \label{eq:channel_integer2}
    \tilde{\bm{v}}_{1,j}^\dagger\bm{y}_1 = 
    \begin{cases} 
        \hat{h}_{1,1} \norm{\bm{v}_{1,j}} (s_{1,j} + s_{2,j})
        +\mu_{1,j}, & \text{if $j \neq L$} \\ 
        \hat{h}_{1,1} \norm{\bm{v}_{1,L}} s_{1,L}
        +\mu_{1,L}, & \text{if $j = L$}
    \end{cases} 
\end{equation} 
where
\begin{equation}
    \label{eq:residualnoise1}
    \mu_{1,j} \defeq  
    \begin{cases}
    e_{1,1,j} s_{1,j} + e_{1,2,j} s_{2,j}+ \theta_{1,j} + \tilde{z}_{1,j},  & \text{if $j\neq L$}\\
    e_{1,1,j} s_{1,j} + \theta_{1,j} + \tilde{z}_{1,j}, & \text{if $j = L$}
    \end{cases}
\end{equation}
is the sum of the imperfect alignment, imperfect zero forcing, and noise
terms.\footnote{The noise term $\mu_{1,j}$ depends on the signal
$s_{k,\ell}$ and is, therefore, not additive. We will handle this
difficulty later.} The signal-to-interference-and-noise ratio (SINR) for
each subchannel at receiver one is thus lower bounded by 
\begin{align}
    \label{eq:sinrlower1}
    \SINR_1 
    & \stackrel{(a)}{\geq} \frac{L\tilde{P}/c^2}{1 + \sigma^2 + 2 \gamma^2 \tilde{P}} \nonumber\\
    & \stackrel{(b)}{\geq} \frac{  L \tilde{P}/c^2}{1 + 2 L \gamma^2 \tilde{P}} \nonumber\\
    & \stackrel{(c)} = \frac{P/(4c^4)}{1 + 2 L^2 (\pi + 1)^2 P/\nu^2 },
\end{align} 
where $(a)$ follows from  \eqref{eq:power}, \eqref{eq:gamma}, \eqref{eq:sigmadef}, 
and \eqref{eq:noise}; $(b)$ follows from \eqref{eq:sigmabound};
and $(c)$ follows from \eqref{eq:psub} and \eqref{eq:residual}.
Similarly, at receiver two, we have 
\begin{equation}
    \label{eq:channel_integer3} 
    \tilde{\bm{v}}_{2,j}^\dagger\bm{y}_2 
    =  \hat{h}_{2,2}\norm{\bm{v}_{2,j}} (b_{j+1}s_{1,j+1} + s_{2,j}) 
    +\mu_{2,j}  
\end{equation}
for $j\in\{1,\dots,L-1\}$ and with
\begin{equation}
    \label{eq:residualnoise2}
    \mu_{2,j} 
    \defeq e_{2,1,j} s_{1,j+1} 
    + e_{2,2,j} s_{2,j} + \theta_{2,j} + \tilde{z}_{2,j}. 
\end{equation} 
Recall that $b_{j+1}$ is of the form $n$ or $1/n$ for some natural
number $n\in\N$ with $n \leq 2c$ by \eqref{eq:bbound}. If $b_{j+1} = n$,
then both channels have integer coefficients.  If $b_{j+1} = 1/n$, then
we can multiply the channel output by $n$ to obtain a channel with
integer coefficients. This decreases the effective SINR by at most a
factor $4c^2$. Following the same steps as before,
the signal-to-interference-and-noise ratio is lower bounded by
\begin{align}
    \label{eq:sinrlower2}
    \SINR_2 
    & \geq \frac{ P/(16c^6)}{1 + 2 L^2 (\pi + 1)^2 P/\nu^2 }. 
\end{align}

As we had seen earlier, the $b_j$ factor serves as a normalizing term to
ensure that all the transmit vectors $\bm{v}_{k,\ell}$ have
approximately magnitude $\sqrt{L}$. From \eqref{eq:channel_integer3}, it
is now clear why $b_j$ has to be chosen as a small integer or its
inverse. Indeed, it is precisely this property that ensures that the
subchannels induced by the matching of channel gains and the
precoder/matched filter have essentially integer channel gains. As we
will see, having integer channel gains significantly simplifies the task
of efficient reliable computation. This transformation of the original
channel with complex coefficients into subchannels with integer
coefficients is at the heart of the proposed communication scheme.

\subsection{Computation of Functions}
\label{sec:proofs_K2_computation}

In the last section, we constructed and analyzed the subchannels induced
by the precoder and matched filter. We now show how to reliably compute
functions over these subchannels from the precoder input to the matched
filter output.

Consider all time slots in the first subblock with quantized channel
realization $\hat{\bm{\msf{H}}}\in\hat{\mc{H}}^{K\times K}$. 
By Lemma~\ref{thm:typical}, with probability at least
$1-\varepsilon_1(T)$ there are at least
\begin{equation}
    \label{eq:that}
    T^{(\hat{\bm{\msf{H}}})}
    \defeq (1-\eta(T)) p_{\hat{\bm{H}}}(\hat{\bm{\msf{H}}})T/L
\end{equation}
time slots in the first subblock that have this quantized channel
realization. By the matching construction in
Section~\ref{sec:proofs_K2_matching}, the first
$T^{(\hat{\bm{\msf{H}}})}$ such time slots can be successfully
matched with time slots in subblocks $\ell \in\{2,\ldots, L\}$ with
quantized channel realizations chosen according to
\eqref{eq:match}. 

By \eqref{eq:channel_integer2} and
\eqref{eq:channel_integer3}, the precoding and matched
filtering scheme from Section \ref{sec:proofs_K2_precoding} transforms
each group of $L$ time slots into $L-1$ subchannels of the form
\begin{subequations}
    \label{eq:subchannel1}
    \begin{align}
        r_{1,j}^{(\hat{\bm{\msf{H}}})}[t] 
        & = \beta_{1,j}^{(\hat{\bm{\msf{H}}})}
        \big(  s_{1,j}^{(\hat{\bm{\msf{H}}})}[t] + s_{2,j}^{(\hat{\bm{\msf{H}}})}[t] \big)
        + \mu_{1,j}^{(\hat{\bm{\msf{H}}})}[t] \\
        r_{2,j}^{(\hat{\bm{\msf{H}}})}[t] 
        & = \beta_{2,j}^{(\hat{\bm{\msf{H}}})}
        \big( a_{1,j+1}^{(\hat{\bm{\msf{H}}})} s_{1,j+1}^{(\hat{\bm{\msf{H}}})}[t] 
        + a_{2,j}^{(\hat{\bm{\msf{H}}})} s_{2,j}^{(\hat{\bm{\msf{H}}})}[t] \big)
        + \mu_{2,j}^{(\hat{\bm{\msf{H}}})}[t]
    \end{align} 
    for $j\in\{1,\ldots, L-1\}$, and where $s_{k,j}^{(\hat{\bm{\msf{H}}})}$ are the channel inputs,
    $a_{1,j+1}^{(\hat{\bm{\msf{H}}})}$ and
    $a_{2,j}^{(\hat{\bm{\msf{H}}})}$ are nonzero integers,
    $\beta_{m,j}^{(\hat{\bm{\msf{H}}})}$ are positive scaling factors,  and
    $\mu_{1,j}^{(\hat{\bm{\msf{H}}})}[t]$ and
    $\mu_{2,j}^{(\hat{\bm{\msf{H}}})}[t]$ are interference and noise
    as in \eqref{eq:residualnoise1} and \eqref{eq:residualnoise2}.
    Receiver one observes one additional subchannel of the form
    \begin{equation}
        r_{1,L}^{(\hat{\bm{\msf{H}}})}[t] 
        = \beta_{1,L}^{(\hat{\bm{\msf{H}}})} s_{1,L}^{(\hat{\bm{\msf{H}}})}[t] 
        + \mu_{1,L}^{(\hat{\bm{\msf{H}}})}[t].
    \end{equation}
\end{subequations}
From \eqref{eq:sinrlower1} and \eqref{eq:sinrlower2}, the $\SINR$ to all
of these subchannels is lower bounded by
\begin{align}
    \label{eq:sinrlower3}
    \SINR({\hat{\bm{\msf{H}}}})
    & \defeq \min_{m} \SINR_m(\hat{\bm{\msf{H}}}) \nonumber\\
    & \geq \frac{ P/(16c^6(\hat{\bm{\msf{H}}}))}{1 + 2 L^2 (\pi + 1)^2 P/\nu^2 }. 
\end{align} 
where we have explicitly written out the dependence of $c$ and 
$\SINR$ on $\hat{\bm{\msf{H}}}$.

Each transmitter $k$ splits its message $w_k$ into non-overlapping
submessages $\bm{w}_{k,j}^{\hat{\bm{\msf{H}}}}$, one for each
subchannel $j$ of quantized channel realization $\hat{\bm{\msf{H}}}$.
Each such submessage is a vector with components in
$\{0,1,\ldots,q-1\}$.  Receiver one attempts to recover the functions 
\begin{align*}
    \bm{u}_{1,j}^{(\hat{\bm{\msf{H}}})}
    \defeq
    \begin{cases}
        \bm{w}_{1,j}^{(\hat{\bm{\msf{H}}})} 
        + \bm{w}_{2,j}^{(\hat{\bm{\msf{H}}})} \pmod{q},
        & \text{if $j\neq L$} \\
        \bm{w}_{1,L}^{(\hat{\bm{\msf{H}}})},
        & \text{if $j=L$}
    \end{cases}
\end{align*} 
over subchannel $j\in\{1,\ldots, L\}$. Receiver two attempts to recover 
the functions
\begin{equation*}
    \bm{u}_{2,j}^{(\hat{\bm{\msf{H}}})} 
    \defeq a_{1,j+1}^{(\hat{\bm{\msf{H}}})}
    \bm{w}_{1,j+1}^{(\hat{\bm{\msf{H}}})} 
    + a_{2,j}^{(\hat{\bm{\msf{H}}})} \bm{w}_{2,j}^{(\hat{\bm{\msf{H}}})} \pmod{q}
\end{equation*}
over subchannel $j\in\{1,\ldots, L-1\}$.

These equations are clearly invertible. Indeed, receiver one decodes
$\bm{w}_{1,L}^{(\hat{\bm{\msf{H}}})}$ alone. Receiver two computes a
linear combination with nonzero coefficients of
$\bm{w}_{2,L-1}^{(\hat{\bm{\msf{H}}})}$ and
$\bm{w}_{1,L}^{(\hat{\bm{\msf{H}}})}$. Knowing
$\bm{w}_{1,L}^{(\hat{\bm{\msf{H}}})}$, we can thus recover
$\bm{w}_{2,L-1}^{(\hat{\bm{\msf{H}}})}$. Continuing in the same manner,
alternating between the receivers in each step, we can successively
recover all transmitted messages. This shows that the mapping between
the messages at the transmitters and the decoded functions at the
receivers is invertible.

Fix a quantized channel realization $\hat{\bm{\msf{H}}}$. Applying $L$
times\footnote{Since the input symbols at the two receivers for
different values of $j\in\{1,\ldots, L\}$ are coupled, we need to make
use of the universality of the channel encoders mentioned after the
statement of Lemma~\ref{thm:lattice}.} \cite[Theorem~1]{nazer11a}
(summarized in the notation of this paper as Lemma~\ref{thm:lattice} in
Appendix~\ref{sec:appendix_lattice}) guarantees that over the
subchannel~\eqref{eq:subchannel1}, a computation sum rate (normalized by
the number $T^{(\hat{\bm{\msf{H}}})}$ of time slots in the subchannel)
arbitrarily close to 
\begin{equation*}
    (2L-1) \log\big(\SINR({\hat{\bm{\msf{H}}}})\big)
\end{equation*} 
is achievable with average probability of error at most
$\varepsilon_2^{(\hat{\bm{\msf{H}}})}(T^{(\hat{\bm{\msf{H}}})}) \to 0$ as
$T^{(\hat{\bm{\msf{H}}})}\to\infty$.  In terms of the original
blocklength $T$, this translates to a computation sum rate of 
\begin{equation*}
    (2L-1)\frac{T^{(\hat{\bm{\msf{H}}})}}{T} 
    \log\big(\SINR({\hat{\bm{\msf{H}}}})\big).
\end{equation*} 
Moreover, since $T^{(\hat{\bm{\msf{H}}})}\to\infty$ as $T\to\infty$, and
since, for fixed $L$ and quantization parameter $\nu$ there are only
finitely many values of $\hat{\bm{\msf{H}}}$, we also have 
\begin{equation*}
    \varepsilon_2(T) 
    \defeq
    \max_{\hat{\bm{\msf{H}}}} 
    \varepsilon_2^{(\hat{\bm{\msf{H}}})}(T^{(\hat{\bm{\msf{H}}})})
    \to 0
\end{equation*}
as $T\to\infty$. 

We repeat the coding procedure above for all quantized channel
realizations $\hat{\bm{\msf{H}}}$ with finite magnitudes, i.e.,
satisfying $\norm{\hat{\bm{\msf{H}}}}_{\infty} < \infty$. If our
construction is successful (see the analysis of error in the following
paragraph), then the overall computation sum rate can be lower bounded
as
\begin{align*}
    (2L-1) & \sum_{\hat{\bm{\msf{H}}}:\norm{\hat{\bm{\msf{H}}}}_{\infty} < \infty} 
    \frac{T^{(\hat{\bm{\msf{H}}})}}{T} \log\big(\SINR({\hat{\bm{\msf{H}}}})\big) \nonumber\\
    & \stackrel{(a)}{\geq} \frac{2L-1}{L} (1-\eta(T))
    \sum_{\hat{\bm{\msf{H}}}:\norm{\hat{\bm{\msf{H}}}}_{\infty} < \infty}
    p_{\hat{\bm{H}}}(\hat{\bm{\msf{H}}}) \log\big(\SINR({\hat{\bm{\msf{H}}}})\big) \nonumber\\
    & \stackrel{(b)}{\geq} \frac{2L-1}{L} (1-\eta(T))
    \sum_{\hat{\bm{\msf{H}}}:\norm{\hat{\bm{\msf{H}}}}_{\infty} < \infty}
    p_{\hat{\bm{H}}}(\hat{\bm{\msf{H}}})
    \bigg( \log\Big( \frac{ P/16}{1 + 2 L^2 (\pi + 1)^2 P/\nu^2 } \Big) 
    - 6\log(c(\hat{\bm{\msf{H}}}))\bigg) \nonumber\\
    & \stackrel{(c)}{\geq} \frac{2L-1}{L} (1-\eta(T))
    \bigg((1- \delta(\nu)) \log\Big( \frac{ P/16}{1 + 2 L^2 (\pi + 1)^2 P/\nu^2 } \Big) 
    - 6\E\big(\log(c(\hat{\bm{\msf{H}}})) ; 
    \norm{\hat{\bm{\msf{H}}}}_{\infty} < \infty\big) \bigg),
\end{align*} 
where $(a)$ follows from \eqref{eq:that}, $(b)$ follows from
\eqref{eq:sinrlower3}, and $(c)$ follows from \eqref{eq:infty}.  Here,
the $(1-\eta(T))$ factor accounts for the loss in matching the channel
gains at times $t_1, \ldots, t_L$, and the factor $(1-\delta(\nu))$
accounts for channel realizations that are quantized to $\infty$, see
Section~\ref{sec:proofs_K2_matching}.  Both $\eta(T) \to 0$ as the
blocklength $T\to\infty$ by \eqref{eq:eta} and $\delta(\nu) \to 0$ as
the quantization parameter $\nu\to\infty$ by \eqref{eq:delta}.  

There are two sources of error in this communication scheme: atypicality
of the channel gains and atypicality of the noise terms. The channel
gains are handled by the matching construction described in
Section~\ref{sec:proofs_K2_matching}.  We declare an error whenever the
channel gains are atypical, which happens with probability at most
$\varepsilon_1(T)$ with $\varepsilon_1(T)\to 0$ as $T\to\infty$ for
fixed $L$ and $\nu$ by \eqref{eq:error} and \eqref{eq:epsilon1}. The
noise is handled by the computation code over the integer channel. As we
have seen above, an error occurs with probability at most
$\varepsilon_2(T)$ with $\varepsilon_2(T) \to 0$ as $T\to\infty$ for
fixed $L$ and $\nu$. Since the number of finite quantized channel gains
is at most $\nu^4 L$ by \eqref{eq:hcard}, and since the number of
decoders is $2L-1\leq 2L$ for each such realization of the quantized
channel, with probability at least
\begin{equation*}
    1- \varepsilon_1(T) -2\nu^4 L^2\varepsilon_2(T)
\end{equation*}
all decoders are successful. For a fixed number of subblocks $L$ and
fixed quantization parameter $\nu$, this quantity converges to one as
$T\to\infty$, yielding an achievable computation sum rate of 
\begin{equation*}
    R(P, L, \nu) 
    \defeq \frac{2L-1}{L} \bigg( (1-\delta(\nu)) \log\bigg(
    \frac{P/16}{1 + 2L^2 (\pi+1)^2P/\nu}
    \bigg)  
    - 6\E\big(
    \log(c(\hat{\bm{H}})); 
    \norm{\hat{\bm{H}}}_\infty < \infty \big)
    \bigg).
\end{equation*}
Hence the computation capacity $C(P)$ is lower bounded as
\begin{equation*}
    C(P) \geq R(P, L, \nu).
\end{equation*}

Since this is true for all values of $\nu$, we may take the limit as
$\nu\to\infty$ to obtain
\begin{align*}
    C(P) 
    & \geq \lim_{\nu\to\infty}R(P, L, \nu) \nonumber \\
    & = 
    \frac{2L-1}{L}\bigg( \log(P/16) 
    - 6\lim_{\nu\to\infty}\E\big( \log(c(\hat{\bm{H}}));
    \norm{\hat{\bm{H}}}_\infty < \infty \big)
    \bigg).
\end{align*}
In Appendix \ref{sec:appendix_fading}, we show that
\begin{equation*}
    \lim_{\nu\to\infty}\E\big( \log(c(\hat{\bm{H}}));
    \norm{\hat{\bm{H}}}_\infty < \infty \big) \leq 3.
\end{equation*}
Thus, the computation capacity is lower bounded by 
\begin{align*}
    C(P)
    & \geq
    \lim_{\nu\to\infty}R(P, L, \nu) \\
    & = \frac{2L-1}{L}\big( \log(P) - 22 \big).
\end{align*}
Finally, we may take a limit as $L\to\infty$, yielding a computation
rate of 
\begin{align*}
    C(P) 
    & \geq \lim_{L\to\infty} \lim_{\nu\to\infty} R(P,L,\nu) \nonumber\\
    & = 2\log(P)-44 \\
    & \geq K\log(P)-7K^3,
\end{align*}
concluding the proof of the lower bound in Theorem~\ref{thm:compute} for
$K=2$. \hfill\IEEEQED

\section{Proof of Lower Bound in Theorem~\ref{thm:compute} for $K>2$ Users}
\label{sec:proofs_K3}

As in the two-user case in Section~\ref{sec:proofs_K2}, the proof for
$K>2$ proceeds in three steps: matching of channel gains (see
Section~\ref{sec:proofs_K3_matching}), linear precoding and matched
filtering (see Section~\ref{sec:proofs_K3_precoding}), and computation
of functions of the messages over the resulting channel from the
precoder input to the matched filter output (see Section
\ref{sec:proofs_K3_computation}). We again quantize all channel gains as
described in Section~\ref{sec:proofs_quantization} and consider large
blocklengths $T$ such that this quantization can be performed for
arbitrarily large quantization parameter $\nu$ and such that the
resulting observed sequence of quantized channel gains is $\eta$-typical
with high probability. Since the effects of quantization and atypicality
are essentially identical to the two-user case, we will not repeat this
analysis here and instead assume directly that $\nu\approx\infty$, which
implies that $\hat{h}_{m,k}[t]\approx h_{m,k}[t]$. The quantization and
typicality arguments for $K=2$ carry over for $K> 2$.

\subsection{Matching of Channel Gains}
\label{sec:proofs_K3_matching}

Fix a large blocklength $T$ and a natural number $I$. Define
\begin{equation*}
    L \defeq (I+1)^{K^2},
\end{equation*}
and divide the block of $T$ channel realizations into $L$ subblocks of
length $T/L$ (assumed to be integer). Consider the channel gains at time
$t_1=1$ in the first of these blocks and the corresponding channel gains
$\bm{H}[t_1]$. Let $t_{\ell}$ be the first time in block $\ell$ such
that\footnote{The probability of this event happening is, of course,
zero. The statement is to be understood in terms of the quantized
channel gains $\hat{h}_{m,k}[t]$ and sufficiently large $\nu$ so that
$\hat{h}_{m,k}[t]\approx h_{m,k}[t]$.}
\begin{equation*}
    h_{m,k}[t_\ell] = \omega_{L}^{(\ell-1)d_{m,k}}h_{m,k}[t_1]
\end{equation*}
for all $k,m\in\{1,\ldots, K\}$, where $\omega_L$ is the
$L$th root of unity as before, and where
\begin{equation*}
    d_{m,k} \defeq (I+1)^{(k-1)K+m-1}.
\end{equation*}
Repeat this construction with $t_1=2$ and so on, ensuring that no time
slot is matched more than once. 

By the assumptions of circular symmetry and ergodicity of the fading
gains, essentially all but a $o(1)$ fraction of the channel gains can be
matched in this fashion as $T\to\infty$ (see
Lemmas~\ref{thm:symmetry} and~\ref{thm:typical}), and we will assume in
the following that $T$ is large enough to ignore the $o(1)$ term (see
Section~\ref{sec:proofs_K2_matching} for a detailed analysis).

\subsection{Precoding and Matched Filtering}
\label{sec:proofs_K3_precoding}

Consider now one such sequence of matched time slots $t_1, \ldots,
t_L$. As in the two-user case, we use linear precoders and matched
filters over the vector channel induced by these $L$ time slots. 
Define the diagonal matrix
\begin{equation*}
    \bm{D}_{m,k} \defeq \diag\big((h_{m,k}[t_\ell])_{\ell=1}^L\big)
\end{equation*}
corresponding to the vector channel of length $L$ between transmitter
$k$ and receiver $m$ at time slots $t_1, \ldots, t_L$.  By construction,
\begin{equation*}
    \bm{D}_{m,k} = h_{m,k}\bm{F}^{d_{m,k}},
\end{equation*}
where 
\begin{equation*}
    h_{m,k} \defeq h_{m,k}[t_1]
\end{equation*}
and 
\begin{equation*}
    \bm{F} \defeq \diag\big((\omega_L^{\ell-1})_{\ell=1}^L\big).
\end{equation*}

Each transmitter uses again a linear precoder with transmit vectors
$\bm{v}\in\mc{V}\subset\C^L$. The set $\mc{V}$ is constructed
as\footnote{This construction of $\mc{V}$ is reminiscent of the one in
\cite[Appendix~III]{cadambe08} for the $K$-user interference channel
with more than three users.}
\begin{equation*}
    \mc{V} 
    \defeq \bigg\{\bigg(\prod_{m,k}
    \Big(\prod_{\alpha=1}^{\alpha_{m,k}}b_{m,k}^{(\alpha)}\Big)
    \bm{D}_{m,k}^{\alpha_{m,k}}\bigg)\bm{1}:
    \alpha_{m,k}\in\{0, \ldots, I-1\}\bigg\}.
\end{equation*}
Since all channel matrices $\bm{D}_{m,k}$ are diagonal by construction,
the product $\bm{D}_{m,k}\bm{D}_{\tilde{m},\tilde{k}}$ commutes, and
hence it is immaterial in which order the product in the definition of
$\mc{V}$ is taken. The scalars $b_{m,k}^{(\alpha)}$ are constructed
recursively, starting from $b_{m,k}^{(1)}$. Each $b_{m,k}^{(\alpha)}$ is
of the form $n$ or $1/n$ for the smallest natural number $n\in\N$ such
that 
\begin{equation*}
    \Big(\prod_{\alpha=1}^{\alpha_{m,k}}b_{m,k}^{(\alpha)}\Big)
    \abs{h_{m,k}}^{\alpha_{m,k}}\in[1,2).
\end{equation*}

As in the two-user case, the role of the
$b_{m,k}^{(\alpha)}$ is to ensure that the transmit vectors all
have approximately the same norm. In particular, 
\begin{equation}
    \label{eq:vbound}
    \sqrt{L} \leq \norm{\bm{v}} \leq 2^{K^2}\sqrt{L}
\end{equation}
for every $\bm{v}\in\mc{V}$. Moreover, by the recursive construction,
\begin{equation}
    \label{eq:bbound2a}
    \big(2\max\big\{\abs{h_{m,k}}, \abs{h_{m,k}}^{-1}\big\}\big)^{-1} 
    \leq \min\big\{b_{m,k}^{(\alpha)}, 1/b_{m,k}^{(\alpha)}\big\} 
    \leq \max\big\{b_{m,k}^{(\alpha)}, 1/b_{m,k}^{(\alpha)}\big\} 
    \leq 2\max\big\{\abs{h_{m,k}}, \abs{h_{m,k}}^{-1}\big\},
\end{equation}
and hence
\begin{equation}
    \label{eq:bbound2}
    \big(2^{K^2}c\big)^{-1}
    \leq \prod_{m,k}\min\big\{b_{m,k}^{(\alpha_{m,k})}, 1/b_{m,k}^{(\alpha_{m,k})}\big\} 
    \leq \prod_{m,k}\max\big\{b_{m,k}^{(\alpha_{m,k})}, 1/b_{m,k}^{(\alpha_{m,k})}\big\} 
    \leq 2^{K^2}c
\end{equation}
for all $\alpha_{m,k}\in\{0,\ldots, I-1\}$, and where
\begin{equation}
    \label{eq:cdef2}
    c 
    = c(\bm{H}) 
    \defeq \prod_{m,k}\max\big\{\abs{h_{m,k}}, \abs{h_{m,k}}^{-1}\big\}.
\end{equation}

Observe that, as in the two-user case, each transmit
vector $\bm{v}\in\mc{V}$ is of the form 
\begin{equation*}
    \bm{v} = \rho\bm{F}^\alpha \bm{1}
\end{equation*}
for some scalars $\rho\in\C$ and $\alpha\in\N$. By the properties of the
``Fourier'' matrix $\bm{F}$, this implies that any two transmit vectors
in $\mc{V}$ are either collinear or orthogonal. As we will see next, all
vectors in $\mc{V}$ are, in fact, orthogonal.

Each $\bm{v}\in\mc{V}$ is a complex-valued vector of length $L$ defined
by a monomial up to power $I-1$ in the channel matrices $\bm{D}_{m,k}$.
By definition, every collection of powers $\alpha_{m,k}\in\{0, \ldots,
I-1\}, m,k\in\{1,\ldots, K\}$ corresponds to an element
$\bm{v}\in\mc{V}$.  We now argue that this correspondence is one-to-one,
implying that
\begin{equation*}
    \card{\mc{V}} = I^{K^2}.
\end{equation*}
Moreover, together with the argument in the last paragraph, this will
also ensure that all vectors in $\mc{V}$ are orthogonal.

To this end, consider $\bm{v}\in\mc{V}$ and write it as
\begin{equation*}
    \bm{v}
    = \rho\Big(\prod_{m,k} \bm{F}^{d_{m,k} \alpha_{m,k}}\Big)\bm{1}
\end{equation*}
for some $\alpha_{m,k}\in\{0, \ldots, I-1\}$ and some scalar $\rho$.
The first component of $\bm{v}$ is equal to $\rho$.  The second
component of $\bm{v}$ is equal to $\rho\omega_L^\alpha$ with 
\begin{align*}
    \alpha 
    & \defeq \sum_{m,k} \alpha_{m,k}d_{m,k} \pmod{L}\\
    & = \sum_{m,k} \alpha_{m,k}(I+1)^{(k-1)K+m-1} \pmod{L}.
\end{align*}
Since each $\alpha_{m,k}\in\{0, \ldots, I-1\}$, this last sum is less
than $(I+1)^{K^2}=L$, and so the modulo $L$ operation can be dropped.
Thus, the coefficients $\alpha_{m,k}$ of $\alpha$ can be determined
uniquely by computing the $(I+1)$-ary expansion of $\alpha$. Moreover,
knowing $\rho$ from the first component of $\bm{v}$, $\alpha$ can be
uniquely determined from the second component of $\bm{v}$. Together,
this shows that there is a unique collection of powers
$\alpha_{m,k}\in\{0, \ldots, I-1\}$ for all $m,k\in\{1,\ldots, K\}$ that
generates $\bm{v}$. We refer to this as the \emph{unique factorization}
property of $\mc{V}$. Since each exponent $\alpha$ corresponds to a
unique $\bm{v}\in\mc{V}$, this also shows the orthogonality of the
vectors in $\mc{V}$.

Each transmitter modulates $I^{K^2}$ zero mean and mutually independent
message symbols over its transmit vectors. Let $s_{k,\bm{v}}$ be the
message symbol at transmitter $k$ sent along transmit vector
$\bm{v}\in\mc{V}$.  The channel input 
\begin{equation*}
    \bm{x}_k \defeq
    \begin{pmatrix}
        x_k[t_1] & x_k[t_2] & \dots & x_k[t_L]
    \end{pmatrix}^\T
\end{equation*}
at transmitter $k$ has then the form
\begin{equation*}
    \bm{x}_k = \sum_{\bm{v}\in\mc{V}} s_{k,\bm{v}}\bm{v}.
\end{equation*}
We allocate the same power 
\begin{equation}
    \label{eq:psub2}
    \E\big(\abs{s_{k,\bm{v}}}^2\big)
    = \frac{P}{4^{K^2}L}
    \defeq \tilde{P}
\end{equation}
to each $s_{k,\bm{v}}$.  Since each transmit vector $\bm{v}$ has
squared norm at most $4^{K^2}L$ by \eqref{eq:vbound}, we have
\begin{equation*}
    \frac{1}{L}\E\big(\norm{\bm{x}_k}^2\big)
    \leq \frac{\card{\mc{V}}}{L}\cdot \frac{P}{4^{K^2}L}\cdot 4^{K^2}L
    \leq P,
\end{equation*}
satisfying the average power constraint over the $L$ time slots
$t_1,\ldots, t_L$. Since each of the $K$ transmitters has $I^{K^2}$
transmit vectors, we transmit a total of $K I^{K^2}$ independent
data streams over $L=(I+1)^{K^2}$ channel uses. 

The corresponding vector of channel outputs
\begin{equation*}
    \bm{y}_m \defeq
    \begin{pmatrix}
        y_m[t_1] & y_m[t_2] & \dots & y_m[t_L]
    \end{pmatrix}^\T
\end{equation*}
at receiver $m$ is then
\begin{align}
    \label{eq:receivechan2}
    \bm{y}_m & = \sum_{k=1}^K \bm{D}_{m,k}\bm{x}_k
    +\bm{z}_m \nonumber\\
    & = \sum_{k=1}^K \sum_{\bm{v}\in\mc{V}} s_{k,\bm{v}}
    \bm{D}_{m,k}\bm{v}
    +\bm{z}_m,
\end{align}
where 
\begin{equation*}
    \bm{z}_m \defeq
    \begin{pmatrix}
        z_m[t_1] & z_m[t_2] & \dots & z_m[t_L]
    \end{pmatrix}^\T
\end{equation*}
is the additive noise at receiver $m$.

From \eqref{eq:receivechan2}, transmit vector $\bm{v}\in\mc{V}$ is
observed at receiver $m$ as $\bm{D}_{m,k}\bm{v}$.  Each receiver $m$
uses $L$ the receive vectors
\begin{equation*}
    \tilde{\mc{V}}_m 
    \defeq \big\{\tilde{\bm{v}} = \bm{D}_{m,k}\bm{v}/\norm{\bm{D}_{m,k}\bm{v}}: 
    k\in\{1,\ldots, K\}, \bm{v}\in\mc{V}\big\}
\end{equation*}
as matched filters, computing $\tilde{\bm{v}}^\dagger\bm{y}_m$ for each 
$\tilde{\bm{v}}\in\tilde{\mc{V}}_m$. The number of matched filters is at
most
\begin{equation*}
    \card{\tilde{\mc{V}}_m}
    \leq (I+1)^{K^2}.
\end{equation*}
By the same argument as for $\mc{V}$, it can be shown
that $\tilde{\mc{V}}_m$ also has the unique factorization property. In
other words, to every $\tilde{\bm{v}}\in\tilde{\mc{V}}_m$ corresponds a unique
collection of powers $\alpha_{m,k}\in \{0, \ldots, I\}$ for all 
$m,k\in\{1,\ldots, K\}$ such that
\begin{equation*}
    \tilde{\bm{v}}
    = \frac{1}{\sqrt{L}}\Big(\prod_{m,k} \bm{F}^{\alpha_{m,k}d_{m,k}}\Big)\bm{1}.
\end{equation*}
As for $\mc{V}$, this implies that the vectors in $\tilde{\mc{V}}_m$ are
orthogonal by the properties of the ``Fourier'' matrix $\bm{F}$.

The equivalent channel, consisting of the linear precoder, the wireless
channel, and the matched filters, has $I^{K^2}$ channel inputs at each
transmitter and at most $(I+1)^{K^2}$ channel outputs at each receiver.
Since the matched filters are normalized to have unit norm, each such
subchannel at the receiver is an additive Gaussian noise channel with
unit noise power. We now argue that we have again signal alignment as in
the two-user case.

As pointed out above, the transmit vector $\bm{v}\in\mc{V}$ at
transmitter $k$ is observed at receiver $m$ as $\bm{D}_{m,k}\bm{v}$. By
construction of the set of matched filter vectors $\tilde{\mc{V}}_m$ at
receiver $m$, $\bm{D}_{m,k}\bm{v}$ is a scalar multiple of a vector
$\tilde{\bm{v}}\in\tilde{\mc{V}}_m$.  Since all the vectors in
$\tilde{\mc{V}}_m$ are orthogonal, this implies that the matched
filtering operation $\tilde{\bm{v}}^\dagger\bm{y}_m$ removes all but
those transmit signals which are aligned with $\tilde{\bm{v}}$.

We now analyze the magnitudes of the signals that are observed along one
receive vector $\tilde{\bm{v}}\in\tilde{\mc{V}}_m$ at receiver $m$. By
unique factorization, there exists a unique collection of exponents
$\alpha_{\tilde{m},\tilde{k}}\in\{0, \ldots, I\}$ such that 
\begin{equation*}
    \tilde{\bm{v}}
    = \rho\Big(\prod_{\tilde{m},\tilde{k}} 
    \bm{D}_{\tilde{m},\tilde{k}}^{\alpha_{\tilde{m},\tilde{k}}}\Big)\bm{1}
\end{equation*}
for some scalar $\rho$. Assume a signal modulated over transmit vector
$\bm{v}_k$ at transmitter $k$ is observed along vector $\tilde{\bm{v}}$
at receiver $m$. Note that this is only possible if $\alpha_{m,k}
\in\{1,\ldots, I\}$ and $\alpha_{\tilde{m}, \tilde{k}}\in\{0, \ldots,
I-1\}$ for all $(\tilde{m}, \tilde{k})\neq (m,k)$. The transmit vector
$\bm{v}_k$ is proportional to $\bm{D}_{m,k}^{-1}\tilde{\bm{v}}$, and
hence is equal to
\begin{equation*}
    \bm{v}_k 
    = \Big(\prod_{\alpha=1}^{\alpha_{m,k}-1}b_{m,k}^{(\alpha)}\Big)
    \bm{D}_{m,k}^{\alpha_{m,k}-1}
    \bigg(\prod_{(\tilde{m},\tilde{k})\neq (m,k)} 
    \Big(\prod_{\alpha=1}^{\alpha_{\tilde{m},\tilde{k}}}b_{\tilde{m},\tilde{k}}^{(\alpha)}\Big)
    \bm{D}_{\tilde{m},\tilde{k}}^{\alpha_{\tilde{m},\tilde{k}}}\bigg)\bm{1}.
\end{equation*}
Defining
\begin{align}
    \label{eq:bkdef}
    b & \defeq 
    \frac{\prod_{\tilde{m},\tilde{k}} 
    \prod_{\alpha=1}^{\alpha_{\tilde{m},\tilde{k}}}b_{\tilde{m},\tilde{k}}^{(\alpha)}}
    {\prod_{\tilde{k}} b_{m,\tilde{k}}^{(\alpha_{m,\tilde{k}})}}, \nonumber\\
    \shortintertext{and} \nonumber\\
    b_k & \defeq \prod_{\tilde{k}\neq k} b_{m,\tilde{k}}^{(\alpha_{m,\tilde{k}})},
\end{align}
this allows to write $\bm{v}_k$ in terms of $\tilde{\bm{v}}$ as
\begin{equation}
    \label{eq:vk}
    \bm{v}_k = 
    \frac{b}{\rho}b_k\bm{D}_{m,k}^{-1}\tilde{\bm{v}}
    \in\mc{V}.
\end{equation}

Since the collection of exponents $\alpha_{\tilde{m}, \tilde{k}}$
corresponding to $\tilde{\bm{v}}$ is unique, and by orthogonality of
$\mc{V}$, this implies that there are at most $K$ signals that are
aligned along the same vector $\tilde{\bm{v}}$ at receiver $m$, and they
are all observed with the same common channel gain times a factor $b_k$
depending on the transmitter $k$. Using the orthogonality of the matched
filters and \eqref{eq:vk}, the output of the
matched filter applied to the channel output \eqref{eq:receivechan2} can then
be written as
\begin{align}
    \label{eq:receivechan3}
    \tilde{\bm{v}}^\dagger\bm{y}_m
    & = \sum_{k=1}^K \sum_{\bm{v}\in\mc{V}} s_{k,\bm{v}}
    \tilde{\bm{v}}^\dagger\bm{D}_{m,k}\bm{v}
    +\tilde{\bm{v}}^\dagger\bm{z}_m \nonumber\\
    & = \sum_{k=1}^K s_{k,\bm{v}_k}
    \tilde{\bm{v}}^\dagger\bm{D}_{m,k}\bm{v}_k
    +\tilde{z}_{m,\tilde{\bm{v}}} \nonumber\\
    & = \frac{b}{\rho} \sum_{k=1}^K b_k s_{k,\bm{v}_k} + \tilde{z}_{m,\tilde{\bm{v}}},
\end{align}
where 
\begin{equation*}
    \tilde{z}_{m,\tilde{\bm{v}}} 
    \defeq \tilde{\bm{v}}^\dagger\bm{z}_m
\end{equation*}
is additive circularly-symmetric complex Gaussian noise with mean zero
and variance one, and where $\bm{v}_k$ depends on both the matched
filter $\tilde{\bm{v}}$ and the receiver $m$ (see~\eqref{eq:vk}). We can
interpret \eqref{eq:receivechan3} as a subchannel between the inputs to
the precoder $\bm{v}_k$ at each transmitter $k$ and the output of
matched filter $\tilde{\bm{v}}$ at receiver $m$.

We point out that, similar to the two-user case, not all $K$
transmitters contribute to all matched filter outputs
$\tilde{\bm{v}}^\dagger\bm{y}_m$. Indeed, if $\alpha_{m,k}=0$ in the
unique factorization of $\tilde{\bm{v}}$ at receiver $m$, then there is
no corresponding transmit vector $\bm{v}_{k}$ at transmitter $k$. For
ease of notation, we assume that $s_{k,\bm{v}_k}=0$ in this case, so
that \eqref{eq:receivechan3} is still valid.

We now bound the channel gains in the matched filter output
\eqref{eq:receivechan3}. From \eqref{eq:vk}, we have
\begin{equation*}
    \abs{b/\rho}
    = \frac{\norm{\bm{v}_k}}{b_k\norm{\bm{D}_{m,k}^{-1}\tilde{\bm{v}}}} \\
    = \frac{\abs{h_{m,k}}\norm{\bm{v}_k}}{b_k}.
\end{equation*}
Now,
\begin{align*}
    \abs{h_{m,k}}^{-1}b_k
    \stackrel{(a)}{=} \abs{h_{m,k}}^{-1}
    \prod_{\tilde{k}\neq k} b_{m,\tilde{k}}^{(\alpha_{m,\tilde{k}})}
    \stackrel{(b)}{\leq} 2^Kc,
\end{align*}
where $(a)$ follows from \eqref{eq:bkdef}, and $(b)$ follows from
\eqref{eq:bbound2a} and \eqref{eq:cdef2}.  Together with
\eqref{eq:vbound}, this shows that
\begin{equation}
    \label{eq:brbound}
    \abs{b/\rho} \geq \frac{\sqrt{L}}{2^Kc}
\end{equation}
Moreover, each $b_k$ is a product of at most $K$ scalars, each being
either a natural number or its inverse.

We want to multiply the output of the subchannel \eqref{eq:receivechan3}
by a positive scalar $\tilde{\rho}$ such that $\tilde{\rho}b_k\in\N$ for all $k$.
By the definition of $b_k$ in \eqref{eq:bkdef}, we can choose
\begin{equation*}
    \tilde{\rho} 
    \defeq \prod_{\tilde{k}=1}^K \max\big\{1,1/b_{m,\tilde{k}}^{(\alpha_{m,\tilde{k}})}\big\}.
\end{equation*}
Using \eqref{eq:bbound2a} and \eqref{eq:bbound2}, we thus have 
\begin{equation}
    \label{eq:trhobound}
    \tilde{\rho} \leq 2^K c,
\end{equation}
resulting in a decrease of effective signal power by at most a factor
$4^K c^2$. 

To summarize, the channel \eqref{eq:receivechan3} between the input
$s_{k,\bm{v}_k}$ to the matched filter at transmitter $k$ and the scaled
output of the matched filter $\tilde{\bm{v}}\in\tilde{\mc{V}}$ at
receiver $m$ is of the form
\begin{equation}
    \label{eq:receivechan4}
    r_{m,\tilde{\bm{v}}}
    = \beta_{m,\tilde{\bm{v}}} \sum_{k=1}^K a_k s_{k,\bm{v}_k} 
    + \mu_{m,\tilde{\bm{v}}},
\end{equation}
for nonzero integer channel gains $a_k$, scaled Gaussian noise
$\mu_{m,\tilde{\bm{v}}}$, and positive scaling factors
$\beta_{m,\tilde{\bm{v}}}$. Ignoring the integer gains $a_k$, the
signal-to-noise ratio
\begin{equation*}
    \SNR
    \defeq \min_{k,m,\tilde{\bm{v}}}
    \frac{\E\abs{\beta_{m,\tilde{\bm{v}}}s_{k,\bm{v}_k}}^2}
    {\E\abs{\mu_{m,\tilde{\bm{v}}}}^2}
\end{equation*}
of each component in this subchannel is then lower bounded by 
\begin{align}
    \label{eq:sinrlower4}
    \SNR 
    & \stackrel{(a)}{\geq} \frac{\tilde{P} \abs{b/\rho}^2}{\tilde{\rho}^2} \nonumber\\
    & \stackrel{(b)}{\geq} \frac{P/(4^{K^2}L)\cdot L/(4^Kc^2)}{4^Kc^2} \nonumber\\
    & = \frac{P}{2^{4K+2K^2}c^4},
\end{align}
where $(a)$ follows from \eqref{eq:psub2}, and $(b)$ follows from
\eqref{eq:brbound} and \eqref{eq:trhobound}.

\subsection{Computation of Functions}
\label{sec:proofs_K3_computation}

We use a computation code over the channel from the precoder input to
the matched filter output constructed in the last section. This will
allow us to reliably decode functions of the transmitted messages over
this channel.

As in the proof of the two-user case, we code over several channel uses,
each with the same channel realization $\bm{\msf{H}}$. For each such
$\bm{\msf{H}}$, we are hence dealing with a channel that is constant
across time. Each transmitter $k$ splits its message $w_k$ into
non-overlapping submessages, one for each subchannel
\eqref{eq:receivechan4} between precoder input and matched filter
output, and for each channel realization $\bm{\msf{H}}$. Each such
submessage is again a vector over $\{0, \ldots, q-1\}$ for some $q$. The
decoder aims to compute a modulo-$q$ integer linear equation of these
messages with coefficients $a_k$ as appearing in
\eqref{eq:receivechan4}.

Using the unique factorization property of $\tilde{\mc{V}}$ and the fact
that all coefficients $a_k$ are nonzero, it follows from \cite[Lemma
8]{niesen12} that the functions to be decoded by the receivers can be
inverted. Hence, knowledge of all correctly decoded functions at the
receivers allows recovery of all the messages. 

Applying $L$ times\footnote{As in the two-user case, the input symbols
at the $K$ receivers are coupled. We make again use of the universality
of the channel encoders mentioned after the statement of
Lemma~\ref{thm:lattice}.} Lemma~\ref{thm:lattice} in
Appendix~\ref{sec:appendix_lattice} shows then that each of the
receivers can reliably compute its desired functions over the
channel given by \eqref{eq:receivechan4} at a sum rate at least
\begin{equation*}
    KI^{K^2}\log\bigl(\SNR(\bm{\msf{H}})\bigr) 
    \geq KI^{K^2}\bigl(\log(P)-4K-2K^2-4\log(c(\bm{\msf{H}}))\bigr)
\end{equation*}
for a particular realization $\bm{\msf{H}}$ of the channel gains, and
where we have used \eqref{eq:sinrlower4}, that the number of messages
sent from each transmitter is $\card{\mc{V}} = I^{K^2}$, and that there
are $K$ receivers. Normalizing by the number $(I+1)^{K^2}$ of channel
uses, we can hence achieve a sum rate of at least
\begin{equation*}
    R(P,I) 
    \defeq \frac{KI^{K^2}}{(I+1)^{K^2}} \Big(\log(P)-4K-2K^2
    -4\E\big(\log(c(\bm{H}))\big)\Big)
\end{equation*}
when averaged over all channel realizations.

The computation sum capacity is then lower bounded as
\begin{equation*}
    C(P) 
    \geq R(P,I).
\end{equation*}
Since this holds for all values of $I$, and since the constant $c$ does
not depend on $I$, we may take the limit as $I\to\infty$ to obtain a
computation rate of at least
\begin{align*}
    C(P) 
    & \geq \lim_{I\to\infty}R(P,I) \\
    & = K\log(P)-4K^2-2K^3
    -4K\E\big(\log(c(\bm{H}))\big) \\
    & \geq K\log(P)-7K^3,
\end{align*}
where we have used the upper bound $3K^2/4$ on the expected value of
$\log(c(\bm{H}))$ in Appendix~\ref{sec:appendix_fading}.  This concludes
the proof of the lower bound in Theorem~\ref{thm:compute} for arbitrary
$K\geq 2$.  \hfill\IEEEQED

\section{Proof of Upper Bound in Theorem~\ref{thm:compute}}
\label{sec:proofs_upper}

The proof adapts an argument from \cite[Theorem~4]{niesen11}.
Since the receivers compute an invertible function of the messages, the 
cut-set bound \cite[Theorem 14.10.1]{cover91} applies, showing that
\begin{equation*}
    C(P)
    \leq \sup_{\bm{Q}(\bm{H})} \E\big(\log\det(\bm{I}+\bm{H}\bm{Q}(\bm{H})\bm{H}^\dagger)\big),
\end{equation*}
where the maximization is over all positive semidefinite matrices
$\bm{Q}(\bm{H})$ such that 
\begin{equation*}
    \E\big(\tr(\bm{Q}(\bm{H}))\big)\leq KP. 
\end{equation*}
Using Hadamard's inequality, this can be upper bounded as
\begin{align*}
    \sup_{\bm{Q}(\bm{H})} \E\big( 
    \log\det(\bm{I}+\bm{H}\bm{Q}(\bm{H})\bm{H}^\dagger)\big) 
    & \leq  \sum_{m=1}^K \sup_{\bm{Q}(\bm{H})}
    \E\big(\log(1+\bm{h}_m\bm{Q}(\bm{H})\bm{h}_m^\dagger)\big) \\
    & \leq K \sup_{P(r)} \E\big(\log(1+rP(r))\big),
\end{align*}
where $\bm{h}_m$ denotes the $m$\/th row of $\bm{H}$, where 
\begin{equation*}
    r \defeq \norm{\bm{h}_1}^2,
\end{equation*}
and where the last maximization is over all nonnegative $P(r)$ satisfying
\begin{equation*}
    \E (P(r)) \leq KP.
\end{equation*}

This upper bound on $C(P)$ is maximized by water-filling
\cite{goldsmith97}, yielding
\begin{equation*}
    C(P) \leq K \E\big(\log(1+rP^\star(r))\big)
\end{equation*}
with
\begin{equation*}
    P^\star(r) \defeq \Big(\frac{1}{\mu}-\frac{1}{r}\Big)^+
\end{equation*}
and $\mu$ such that
\begin{equation}
    \label{eq:upper_power}
    \E(P^\star(r)) = KP.
\end{equation}
Since
\begin{equation*}
    P^\star(r) \leq \frac{1}{\mu},
\end{equation*}
we can further upper bound
\begin{align}
    \label{eq:cupper}
     C(P) 
     & \leq K \E(\log(1+r/\mu)) \nonumber\\
     & \leq K \log(1+\E(r)/\mu),
\end{align}
where we have used Jensen's inequality.

It remains to lower bound $\mu$. By \eqref{eq:upper_power}, we have
\begin{align*}
    KP 
    & = \E(P^\star(r)) \\
    & = \int_{\msf{r}=\mu}^\infty \Big(\frac{1}{\mu}-\frac{1}{\msf{r}}\Big)
    f_r(\msf{r})d\msf{r} \\
    & \geq \int_{\msf{r}=2\mu}^\infty \Big(\frac{1}{\mu}-\frac{1}{\msf{r}}\Big)
    f_r(\msf{r})d\msf{r} \\
    & \geq \frac{1}{2\mu}\Pp(r \geq 2\mu).
\end{align*}
The random variable $r$ has Erlang distribution with parameter $K$ and
rate one, and hence
\begin{align*}
    KP 
    & \geq \frac{1}{2\mu}\Pp(r \geq 2\mu) \\
    & = \frac{1}{2\mu}\sum_{k=0}^{K-1} \exp(-2\mu)\frac{(2\mu)^k}{k!} \\
    & \geq \frac{1}{2\mu}\exp(-2\mu).
\end{align*}
If  $\mu \leq 1/(4KP)$, then we obtain the contradiction
\begin{align*}
    KP 
    & \geq \frac{1}{2\mu}\exp(-2\mu)  \\
    & \geq 2KP\exp(-1/(2KP))  \\
    & > KP
\end{align*}
for $K\geq 2$, $P\geq 1$. Hence $\mu > 1/(4KP)$. 

Substituting this into \eqref{eq:cupper} yields
\begin{align*}
    C(P) 
    & \leq K \log(1+4KP \E(r)) \\
    & = K \log(1+4K^2P) \\
    & \leq K\log(P) + 5K\log(K),
\end{align*}
where we have used $P\geq 1$ and $K \geq 2$.  This concludes the proof
of the upper bound in Theorem~\ref{thm:compute}. \hfill\IEEEQED

\section{Proof of Theorem~\ref{thm:multiple}}
\label{sec:proofs_multiple}

This section provides the proof for the approximation result of the sum
capacity $C^{(D)}(P)$ of the $D$-layer relay network. The proof builds
on the approximation result for the computation sum rate in
Theorem~\ref{thm:compute}. Since the upper bound in
Theorem~\ref{thm:multiple} follows directly from the same cut-set bound
argument as Theorem~\ref{thm:compute}, we focus here on the lower bound. 

Each of the $D$ network layers operates using compute-and-forward. We
use the same codebook rate $R_k = R$ at each source node
$k\in\{1,\ldots, K\}$. Using Theorem~\ref{thm:compute}, the relay nodes
at layer one can then reliably decode a deterministic invertible
function of the messages at sum rate at least
\begin{equation*}
    K\log(P)-7K^3.
\end{equation*}
Since the blocklength used is arbitrarily long, the probability of
decoding error at the relays can be made smaller than $\varepsilon/D$
for any $\varepsilon> 0$. 

The relays in layer one treat these computed functions as their messages
for the destination node, and re-encode them using again a computation
code. In order to make this argument inductively, we will apply
Theorem~\ref{thm:compute} for each layer. Two difficulties arise. First,
the statement in Theorem~\ref{thm:compute} is only for the computation
\emph{sum rate} and it is not clear how much each individual transmitter
and receiver contributes to this sum. For the induction argument, we
need to argue that we can choose the message rates at the transmitters
to be symmetric, and that we can choose the rates of the decoded
functions at the receivers to be symmetric. Second, the definition of
computation capacity stipulates only that the receivers decode an
invertible deterministic function of the messages. In particular, the
sum rate of the decoded functions at any receiver could be larger than
the sum rate of the transmitted messages. For example, if a receiver
decodes a sum over $\Z$ of two messages, then the entropy of this
decoded function is larger than the entropy of either of the messages.
For the induction argument, we need to argue that the we can choose the
functions to be computed at the receivers to be over the same alphabet
as the messages at the transmitters, thus avoiding growth of the
messages as they traverse the network.

From the proof of Theorem~\ref{thm:compute}, we see that the rates of
the messages at the transmitters as well as the rates of the computed
functions at the receivers are indeed symmetric as the time expansion
parameter $L\to\infty$ (see Sections~\ref{sec:proofs_K2_computation} and
\ref{sec:proofs_K3_computation}). Moreover, the messages at the
transmitters as well as the computed functions at the receivers are all
over the same finite field of size $q$ (see again
Sections~\ref{sec:proofs_K2_computation} and
\ref{sec:proofs_K3_computation}). Thus, the message sizes do not
increase as they traverse the network.

We can therefore inductively apply Theorem~\ref{thm:compute} to conclude
that the relays at layer $d$ in the network can decode a deterministic
invertible function of the messages at layer $d-1$ for all $d\in\{1,
\ldots, D\}$ at sum rate at least
\begin{equation*}
    K\log(P)-7K^3.
\end{equation*}
Since the composition of invertible functions is invertible, this
implies that the relay nodes in layer $D$ compute a deterministic
invertible function of the messages at the source at this sum rate.

Since the relay nodes in the last layer are connected to the destination
node by orthogonal bit pipes of infinite capacity, they can forward
their computed message to the destination. The destination node, in
turn, can then invert these $K$ functions to recover the original
messages. Since the probability of decoding error is at most
$\varepsilon/D$ in each layer, this implies that the destination node
decodes in error with probability at most $\varepsilon$ by the union
bound. Since $\varepsilon>0$ is arbitrary, this proves the lower bound
in Theorem~\ref{thm:multiple}. \hfill\IEEEQED

\section{Conclusions}
\label{sec:conclusions}

We have considered time-varying Gaussian relay networks consisting of $K$
source nodes communicating to a destination node with the help of $D$
layers of $K$ relay nodes. We have presented a capacity approximation
for this type of communication network. The gap in this approximation
depends only on the number of source nodes $K$ and the fading statistics,
but is independent of the depth $D$ of the network and the transmit
power $P$. This contrasts with previously known approximation results,
which have a gap that increases linearly with the depth $D$ of the
network.

At the heart of our achievable scheme is the concept of computation
alignment, combining computation codes with signal alignment. The use of
computation codes allows the relay nodes to remove receiver noise, thus
preventing noise from accumulating as messages traverse the network. The
use of signal alignment allows the transformation of the wireless
channel with time-varying complex-valued channel gains into subchannels
with constant integer-valued channel gains, over which these computation
codes can be used efficiently.

\appendices

\section{Computation Over Integer Channels}
\label{sec:appendix_lattice}

The channel matching and precoding/matched filtering steps in
Sections~\ref{sec:proofs_K2} and \ref{sec:proofs_K3} transform the
time-varying linear channel with arbitrary complex channel gains into
several constant linear subchannels with integer channel gains. In this
section, we analyze how to reliably compute functions over these
subchannels.  We will employ the compute-and-forward scheme from
\cite{nazer11a}, being well-suited for such constant linear channels
with integer channel gains.

Throughout this section, we consider the subchannels~\eqref{eq:subchannel1}
and \eqref{eq:receivechan4}. Specifically, relay $m$ observes
\begin{equation}
    \label{eq:subchannel2} 
    r_{m}[t] 
    \defeq \beta \sum_{k=1}^K a_{m,k}  s_{k}[t] + \mu_{m}[t] 
\end{equation}
where $\beta >0$ is a positive real scaling factor, $a_{m,k} \in \Z$ are integer
channel coefficients, $s_{k}[t]\in\C$ are the symbols sent by transmitter
$k$, and
\begin{equation*}
    \mu_{m}[t] 
    \defeq \sum_{k=1}^K e_{m,k}[t] s_{k} [t] + \theta_{m}[t] + z_{m}[t]
    \in\C
\end{equation*} 
is the sum of interference and noise terms. Part of the interference is
due to residual channel fluctuations $e_{m,k}[t]$ and the remainder is due
to leakage from other subchannels written as $\theta_{m}[t]$.
We assume that
\begin{equation*}
    \abs{e_{m,k}[t]} \leq \gamma^2
\end{equation*}
for all $m,k$, and for some finite constant $\gamma^2$ not depending on
$m,k,t$. Finally, $z_{m}[t]$ is i.i.d. circularly-symmetric Gaussian
noise with mean zero and variance one. Each leakage term $\theta_{m}[t]$
has expected power 
\begin{equation*}
    \E\big(\abs{\theta_{m}[t]}^2\big) \leq \sigma^2 
\end{equation*}
and is independent of the symbols $s_{k}[t]$ for all $m,k,$ and $t$.
Over a block of length $T$, we impose an average power constraint of
\begin{equation*}
    \frac{1}{T}\sum_{t=1}^T \abs{s_{k}[t]}^2 \leq P.  
\end{equation*} 

It will be convenient to express the messages at the transmitters as
well as the functions computed at the receivers in some finite
field.\footnote{This property will be quite useful in the analysis of
$D$-layer relay networks as it ensures that the rates of the recovered
functions are the same as the transmitted messages.} To this end, we
write the message $w_k$ at transmitter $k$ as a vector $\bm{w}_k$ of
length $\kappa$ with components in $\{0,\ldots, q-1\}$ for some prime
number $q$. Receiver $m$ aims to recover the function
\begin{equation*}
    \bm{u}_m \defeq 
    \sum_{k=1}^K a_{m,k} \bm{w}_k \pmod{q}
\end{equation*} 
where $a_{m,k}$ are the same integer-valued coefficients that appear in
\eqref{eq:subchannel2}. We will assume that these coefficients are
chosen so that the resulting functions are invertible.
Since we transmit $K$ messages with alphabet size $q^\kappa$ over $T$
channel uses, the computation sum rate (in bits per channel use) is
\begin{equation*}
    K \frac{\kappa}{T} \log(q). 
\end{equation*} 

The following result, which is a special case of
\cite[Theorem~1]{nazer11a}, lower bounds the computation sum capacity of
the channel~\eqref{eq:subchannel2}.
\begin{lemma} 
    \label{thm:lattice}
    The computation sum capacity of the channel~\eqref{eq:subchannel2}
    is lower bounded by
    \begin{equation*}
        K \log(\SINR)
    \end{equation*} 
    with
    \begin{equation*}
        \SINR \defeq \frac{\beta^2 P}{1+ \sigma^2 +  K \gamma^2 P}.
    \end{equation*}
\end{lemma}

We point out that the codebooks at the $K$ transmitters in
Lemma~\ref{thm:lattice} are chosen independently of the coefficients
$a_{m,k}$. In other words, the encoders are universal with respect to
the channel and equation coefficients $a_{m,k}$.

\section{Upper Bound on the Expected Value of $\log(c(\hat{\bm{H}}))$}
\label{sec:appendix_fading}

In this section, we derive the upper bound 
\begin{equation*}
    \lim_{\nu\to\infty}\E\big( \log(c(\hat{\bm{H}}));
    \norm{\hat{\bm{H}}}_\infty < \infty \big) \leq \frac{3K^2}{4}
\end{equation*}
as the quantization parameter $\nu\to\infty$. 

The term $c$ depends on the quantized channel gains $\hat{\bm{H}}$, and
hence, implicitly, on the channel gains $\bm{H}$ and the quantization
parameter $\nu$. With slight abuse of notation, we write
\begin{equation*}
    c(\hat{\bm{H}}) = c(\bm{H}, \nu).
\end{equation*}
We then have
\begin{align*}
    \label{eq:c1}
    \E\big( \log( c(\hat{\bm{H}})); \norm{\hat{\bm{H}}}_\infty < \infty \big) 
    & = \sum_{\hat{\bm{\msf{H}}}: \norm{\hat{\bm{\msf{H}}}}_\infty < \infty}
    \log(c(\hat{\bm{\msf{H}}})) p_{\hat{\bm{H}}}(\hat{\bm{\msf{H}}}) \\
    & =\sum_{\hat{\bm{\msf{H}}}: \norm{\hat{\bm{\msf{H}}}}_\infty < \infty}
    \log(c(\hat{\bm{\msf{H}}})) \int_{\bm{\msf{H}}\in Q^{-1}(\hat{\bm{\msf{H}}})}
    f_{\bm{H}}(\bm{\msf{H}})d\bm{\msf{H}} \\
    & = \int_{\bm{\msf{H}}: \norm{\bm{\msf{H}}}_\infty \leq \nu}
    \log(c(\bm{\msf{H}},\nu)) 
    f_{\bm{H}}(\bm{\msf{H}})d\bm{\msf{H}} \\
    & = \E\big( \log(c(\bm{H}, \nu)); \norm{\bm{H}}_\infty\leq \nu\big)
\end{align*}
by Fubini's theorem, and where 
$f_{\bm{H}}$ denotes the density of $\bm{H}$ and $Q$ the operation of
the quantizer.

From the definition of $c$, and using \eqref{eq:quant2},
\begin{align*}
    c(\bm{H}, \nu) 
    & = \prod_{m,k} \max\big\{\abs{\hat{h}_{m,k}}, \abs{\hat{h}_{m,k}}^{-1}\big\}  \\
    & \leq 2^{K^2} \prod_{m,k} \max\big\{\abs{h_{m,k}}, \abs{h_{m,k}}^{-1}\big\}
\end{align*}
for $\bm{H}$ such that $\norm{\bm{H}}_{\infty}\leq \nu$. Hence,
\begin{equation*}
    \log(c(\bm{H}, \nu))
    \ind\{\norm{\bm{H}}_\infty\leq \nu\} \\
    \leq K^2 + \sum_{m,k} 
    \log\big(\max\big\{\abs{h_{m,k}}, \abs{h_{m,k}}^{-1}\big\}\big).
\end{equation*}
Since
\begin{equation*}
    \E \big(\log\big(
    \max\big\{\abs{h_{m,k}}, \abs{h_{m,k}}^{-1}\big\}
    \big) \big)
    < \infty
\end{equation*}
by assumption on the fading process, this implies that
\begin{equation*}
    \lim_{\nu\to\infty}
    \E\big(
    \log(c(\bm{H}, \nu));
    \norm{\bm{H}}_\infty \leq \nu \big)
    = \E\big(\lim_{\nu\to\infty}\log(c(\bm{H}, \nu))\big)
\end{equation*}
by dominated convergence. Since $\hat{\bm{H}}$ converges to $\bm{H}$
almost surely as $\nu\to\infty$ by the construction of the quantizer,
this yields
\begin{align}
    \label{eq:c2}
    \lim_{\nu\to\infty}
    \E\big( 
    \log(c(\bm{H}, \nu));
    \norm{\bm{H}}_\infty \leq \nu \big) 
    & = \sum_{m,k} \E\big(
    \log\big(\max\big\{\abs{h_{m,k}}, \abs{h_{m,k}}^{-1}\big\}\big) \big) \nonumber\\
    & = \frac{K^2}{2}\E \big(
    \log\big(\max\big\{\abs{h_{1,1}}^2, \abs{h_{1,1}}^{-2}\big\}\big)
    \big).
\end{align}

It remains to upper bound the expectation over $h_{1,1}$. Since
$\abs{h_{1,1}}^2$ has exponential distribution, we have
\begin{align*}
    \E\big( \log\big(\max\big\{\abs{h_{1,1}}^2,
    \abs{h_{1,1}}^{-2}\big\}\big) \big) 
    & = -\int_{s=0}^1 \exp(-s)\log(s)ds 
    + \int_{s=1}^\infty \exp(-s)\log(s)ds \nonumber\\
    & = (\gamma - 2 \Ei(-1))\log(e) \nonumber\\
    & \leq 1.5,
\end{align*}
where $\gamma$ is the Euler-Mascheroni constant. 
Combining this with \eqref{eq:c2} shows that
\begin{equation*}
    \lim_{\nu\to\infty}
    \E\big( \log( c(\hat{\bm{H}})); \norm{\hat{\bm{H}}}_\infty < \infty \big) 
    = \E\big(\lim_{\nu\to\infty}\log(c(\bm{H}, \nu))\big)  
    \leq \frac{3K^2}{4}.
\end{equation*}

\section*{Acknowledgment}

The authors would like to thank the reviewers for their thoughtful comments.

\end{document}